\documentclass[a4paper,fleqn,usenatbib]{mnras}

\usepackage{newtxtext,newtxmath}
\usepackage{nicefrac,xfrac}

\usepackage[T1]{fontenc}
\usepackage{ae,aecompl}

\usepackage{graphicx}	
\usepackage{amsmath}	

\title[LOFAR Detection of a Radio Halo in Abell 990]{LOFAR Detection of a Low-Power Radio Halo in the Galaxy Cluster Abell 990}

\author[N. D. Hoang et al.]{N. D. Hoang ,$^{1}$\thanks{E-mail: \href{mailto:hoang@hs.uni-hamburg.de}{hoang@hs.uni-hamburg.de}}
T. W. Shimwell,$^{2,3}$
E. Osinga,$^{3}$
A. Bonafede,$^{4}$
M. Br\"uggen,$^{1}$\newauthor
A. Botteon$^{3}$,
G. Brunetti,$^{5}$
R. Cassano$^{5}$,
V. Cuciti $^{1}$
A. Drabent$^{6}$,
C. Jones$^{7}$,\newauthor
H. J. A. R\"{o}ttgering,$^{3}$
and R. J. van Weeren$^{3}$
\\ \\
$^{1}$Hamburger Sternwarte, University of Hamburg, Gojenbergsweg 112, 21029 Hamburg, Germany\\
$^{2}$Netherlands Institute for Radio Astronomy (ASTRON), P.O. Box 2, 7990 AA Dwingeloo, The Netherlands\\
$^{3}$Leiden Observatory, Leiden University, PO Box 9513, NL-2300 RA Leiden, the Netherlands\\
$^{4}$Department of physics and astronomy, Bologna University - Via Zamboni, 33, 40126 Bologna, Italy\\
$^{5}$IRA INAF, via P. Gobetti 101 40129 Bologna, Italy\\
$^{6}$Th\"uringer Landessternwarte, Sternwarte 5, 07778 Tautenburg, Germany \\
$^{7}$Harvard-Smithsonian Center for Astrophysics, 60 Garden Street, Cambridge, MA 02138, USA
}
\date{Accepted 2020 November 13. Received 2020 October 15; in original form 2020 May 18
}

\pubyear{2020}

\begin{document}
\label{firstpage}
\pagerange{\pageref{firstpage}--\pageref{lastpage}}
\maketitle

\begin{abstract}

Radio halos are extended ($\sim{\rm Mpc}$), steep-spectrum sources found in the central region of dynamically disturbed clusters of galaxies. Only a handful of radio halos have been reported to reside in galaxy clusters with a mass   $M_{500}\lesssim5\times10^{14}\,M_\odot$. In this paper we present a LOFAR 144 MHz detection of a radio halo  in the galaxy cluster Abell 990 with a mass of $M_{500}=(4.9\pm0.3)\times10^{14}\,M_\odot$. The halo has a projected size of $\sim$700$\,{\rm kpc}$ and a flux density of $20.2\pm2.2\,{\rm mJy}$ or a radio power of $1.2\pm0.1\times10^{24}\,{\rm W\,Hz}^{-1}$ at the cluster redshift ($z=0.144$) which makes it one of the two halos with the lowest radio power detected to date. Our analysis of the emission from the cluster with Chandra archival data using dynamical indicators shows that the cluster is not undergoing a major merger but is a slightly disturbed system with a mean temperature of $5\,{\rm keV}$. The low X-ray luminosity of $L_{X}=(3.66\pm0.08)\times10^{44}\,{\rm ergs\,s}^{-1}$ in the  0.1--2.4~keV band implies that the cluster is one of the least luminous systems known to host a radio halo. Our detection of the radio halo in Abell 990 opens the possibility of detecting many more halos in poorly-explored less-massive clusters with low-frequency telescopes such as LOFAR, MWA (Phase II) and uGMRT.

\end{abstract}

\begin{keywords}
galaxies: clusters: individual (Abell 990)
\textendash{} galaxies: clusters: intra-cluster medium \textendash{}
large-scale structure of Universe \textendash{} radiation mechanisms:
non-thermal \textendash{} diffuse radiation

\end{keywords}



\section{Introduction}

Diffuse ($\sim$Mpc) synchrotron radio emission (i.e. halos) in the centre of galaxy clusters indicates the presence of relativistic particles and large-scale magnetic fields in the intra-cluster medium (ICM). It is well known that the acceleration of particles to relativistic energy and the amplification of magnetic fields at these scales are associated with  dynamical events in the host cluster of galaxies. However, the physical mechanisms governing the particle acceleration and the magnetic field amplification have not been fully understood \citep[e.g.,][]{Brunetti2014,VanWeeren2019a}.

Radio halos are extended ($\sim$Mpc), faint ($\sim$$\upmu{\rm Jy\,arcsec}^{-2}$ at 1.4~GHz), steep-spectrum\footnote{the convention $S\propto\nu^\alpha$ is used in this paper} ($\alpha\lesssim-1$) sources that pervade large regions of the cluster. They are apparently unpolarised down to a few percent  above $\sim$1 GHz. In turbulent re-acceleration model \citep[e.g.][]{Brunetti2001,Petrosian2001a,Brunetti2007a}, radio halos are generated by merger-induced turbulence that accelerates cosmic ray (CR) particles and/or amplifies the magnetic fields in the ICM. In addition, hadronic models that attempt to explain the generation of relativistic particles by the inelastic collisions between relativistic protons and thermal ions in the ICM seems to play a lesser role in the formation of radio halos \citep{Ackermann2010,Brunetti2014,Ackermann2016b,Brunetti2017}, but could be a more significant process in mini-halos that have smaller sizes  \citep[$\lesssim500\,{\rm kpc}$; ][]{Pfrommer2004a,Zandanel2014a,Brunetti2014}.

Observational evidence shows that the generation of radio halos depends on the dynamical state of the host clusters and their mass \citep[e.g.][]{Wen2013,Cuciti2015}. Radio halos are more likely to be observed in clusters that are massive and highly disturbed.

The power of radio halos is known to roughly correlate with their cluster mass and X-ray luminosities \citep[e.g.][]{Cassano2013,Birzan2019}. The scatters in the correlations reflect the evolution of halos and dynamical state of clusters \cite[e.g.][]{Cuciti2015}. Spectral properties of the halos are also a contributing factor to the scatters. According to turbulent re-acceleration models these properties reflect the efficiency of particle acceleration and are connected with cluster mass and dynamics. Steep spectrum halos are generally expected in less massive systems and at higher redshift, and are thought to be less powerful sitting below the radio power - cluster mass correlation \cite[e.g.][]{Cassano2006,Cassano2010,Cassano2013,Brunetti2008}.

Our understanding of the formation of radio halos are mainly based on studies of galaxy clusters that are more massive than $5\times10^{14}\,M_{\odot}$. There have been only a handful of radio halos detected in clusters that are less massive than $5\times10^{14}\,M_{\odot}$  (Abell 545, \citealt{Giovannini1999,Bacchi2003}; Abell 141 \citealt{Duchesne2017}; Abell 2061, \citealt{Rudnick2009};  Abell 2811, \citealt{Duchesne2017}; Abell 3562, \citealt{Venturi2001a,Venturi2003,Giacintucci2005}; PSZ1~G018.75+23.57, \citealt{Bernardi2016}; RXC~J1825.3+3026, \citealt{Botteon2019}; Abell 2146, \citealt{Hlavacek-Larrondo2017,Hoang2019b}). This is due to the sensitivity  limitation of previous radio observations that are unable to detect faint radio halos in low-mass clusters. As a consequence, radio halos in the regime of low-mass clusters are largely unexplored. Steep-spectrum nature of radio halos makes low-frequency telescopes such as LOw Frequency ARray (LOFAR; \citealt{VanHaarlem2013}) ideal instruments for studying radio halos in the low-mass regime.

In this paper, we present our search for extended radio emission from the galaxy cluster Abell 990 (hereafter A990). With a mass of $M_{\rm 500}=(4.9\pm0.3)\times10^{14}M_\odot$  \citep[][]{Planck2015}, A990 is an excellent target to search for cluster-scale radio emission in a mass range that is below the typical one where radio halos are found. We make use of the LOFAR Two-metre Sky Survey \citep[LoTSS;][]{Shimwell2017, Shimwell2019}  $120 -168$~MHz  data. The Karl G. Jansky Very Large Array (VLA) $1-2$~GHz data and the archival NRAO VLA Sky Survey (NVSS) 1.4~GHz data are used to constraint the spectral properties the extended radio sources. In addition, we also use the archival X-ray Chandra data (Obs ID: 15114) to study dynamical state of the cluster.

We assume a flat $\Lambda$CDM cosmology with $\Omega_M=0.3$, $\Omega_\Lambda=0.7$, and $H_0=70$~km~s$^{-1}$~Mpc$^{-1}$.  With the adopted cosmology, $1\arcmin$ corresponds to a physical size of $151.6$~kpc at the cluster redshift, i.e. $z=0.144$. The luminosity distance to the cluster is $D_{L}=682.1\,{\rm Mpc}$.

\section{Observations and Data reduction}

\subsection{LOFAR data}
\label{sec:lofar} 

LoTSS is an on-going 120--168~MHz  survey of the entire northern hemisphere \citep[][]{Shimwell2017, Shimwell2019}. As of 4th March 2020, 1522 of 3168 pointings have been observed. The calibrated data covering 27 percent of the northern sky will be released in the upcoming Data Release 2. A990 was observed with LOFAR, as part of LoTSS, during Cycle 4 and is covered by three pointings: P154+50, P156+47 and P158+50. Details of the  observations are given in Table~\ref{tab:obs}.

\begin{table*}
	\centering
	\caption{Observation details}
	\begin{tabular}{lcc}
		\hline\hline
		Telescope                 &    LOFAR 144 MHz$^a$    & VLA 1.5 GHz          \\ \hline
		Project/Pointing              & LC4\_034/P154+50, P156+47, P158+50 &    15A-270          \\
		Configurations            &        HBA\_DUAL\_INNER        &      B             \\
		Observing dates           &    Aug. 08, Jul. 21, Jun. 08,  2015 & Feb. 22, 2015   \\
		Obs. IDs                  &  L345594, L351840, L345592   &     -       \\
		Calibrators               &   3C~196   &   3C~196, 3C~286      \\
		Frequency (MHz)  &      $120-168$      &   $1000-2000$       \\
		Bandwidth (MHz)           &         48          &      1000            \\
		On-source time (hr) &         24          &      0.6         \\
		Integration time (s)      &          1          &         3       \\
		Frequency resolution (kHz) & 12.2  &    1000  \\
		Correlations              &   XX, XY, YX, YY    &    RR,  RL,  LR,  LL   \\
		Number of stations/antennas        &         59, 60, 59          &      26            \\ \hline\hline
	\end{tabular}\\
	Notes: $^a$ for a detail description of the LoTSS data, see \citealt{Shimwell2017,Shimwell2019}. 
	\label{tab:obs}
\end{table*}

The LOFAR data was calibrated in two steps to correct for the direction-independent and direction-dependent effects using   $\mathtt{Prefactor}$\footnote{\url{https://www.astron.nl/citt/prefactor}} \citep[][]{VanWeeren2016a,Williams2016a,DeGasperin2019} and  $\mathtt{ddf-pipeline}$\footnote{\url{https://github.com/mhardcastle/ddf-pipeline}}\citep[][]{Tasse2014a,Tasse2014b,Smirnov2015,Tasse2018,Shimwell2019}, respectively. We briefly outline the procedure below. For a full description of the procedure, we refer to \cite{Shimwell2019}.
	
In the direction-independent calibration, the absolute amplitude is corrected according to the \cite{Scaife2012} flux scale. The calibration solutions are derived from the observations of 3C~196 which model has a total flux density of $83.1\,{\rm Jy}$ at $150\,{\rm MHz}$. The initial XX-YY phase and clock offsets for each station are derived from the amplitude calibrator and are transferred to the target data. The target data are flagged and corrected for ionospheric Faraday rotation\footnote{\url{https://github.com/lofar-astron/RMextract}}. During the data processing, the radio frequency interference (RFI) is removed with $\mathtt{AOFlagger}$ \citep[][]{Offringa2012c} when needed. The data are then phase calibrated against a wide-field sky model that is extracted from the  GMRT 150 MHz All-sky Radio Survey (TGSS-ADR1) catalogue \citep{Intema2017}. 

The direction dependent calibration step mainly aims to correct for the directional phase distortions caused by the ionosphere and the errors in the primary beam model of the LOFAR High Band Antennas (HBA). The direction dependent corrections are solved for each antenna in $\mathtt{kMS}$\footnote{\url{https://github.com/saopicc/killMS}} \citep[][]{Tasse2014a,Tasse2014b,Smirnov2015} implemented in the $\mathtt{ddf-pipeline}$. The flux densities of the LOFAR detected sources are corrected with the bootstrapping technique using the VLSSr and WENSS catalogues  \citep[][]{Hardcastle2016}. Prior to the bootstrapping, the WENSS catalogue is scaled by a factor of 0.9 to be consistent with the flux density scale in LoTSS \citep[i.e.][]{Scaife2012}. To improve the image quality, the pipeline uses  a ``lucky imaging'' technique that generates additional weights to the visibilities basing on the quality of the calibration solutions \citep[][]{Bonnassieux2018}. For imaging, $\mathtt{DDFacet}$ imager \citep{Tasse2018} is used to deconvolve the calibrated data on each facet.

 To improve the quality of the final images in the cluster region, the data are processed through  ``extraction'' and self-calibration steps. In the extraction, all sources outside of the cluster region are subtracted because they are not of the interest of this work. The sources outside of a $40'\times40'$ box centred at the cluster are subtracted using the directional calibration solutions from the $\mathtt{ddf-pipeline}$ run. After the subtraction, data from all pointings are phase-shifted to the cluster position, averaged in time and frequency, and compressed with Dysco \citep{Offringa2016} to reduce data size. The data are corrected for the primary beam attenuation in the direction of the cluster by multiplying with a factor that is inversely proportional to the station beam response. The requirement for the beam correction here is that the region of interest (i.e. the angular size of the cluster) is small enough for the station beam response to be approximately uniform. In addition, the total flux density in the extraction region should be above $\sim$$0.3$~Jy for the calibration to be converged. Details of the ``extraction'' and self-calibration steps are described in \cite{VanWeeren2020}. 

The combined data sets are processed with multiple self-calibration iterations including `tecandphase' and gain calibration rounds. After each iteration, the calibration solutions are smoothed with $\mathtt{LoSoTo}$\footnote{\url{https://github.com/revoltek/losoto}} \citep{DeGasperin2019} to minimize the effect of  noisy solutions. The final intensity images of the cluster at different resolution are created using $\mathtt{WSClean}$\footnote{\url{https://gitlab.com/aroffringa/wsclean}} \citep{Offringa2014} with the imaging parameters in Table~\ref{tab:image_para}. To enhance extended sources, we subtract compact sources from the $uv$ data and make images of the cluster with more weightings on the short $uv$ baselines. The models for compact sources used in the subtraction are obtained from re-imaging the cluster using only data with $uv$ distance longer than $1.0\,k\lambda$, which do not contain extended emission larger than $4.2\arcmin$ (i.e. $637\,{\rm kpc}$ at the cluster redshift). 

To examine the flux scale in the LOFAR images, we compare the flux densities of nearby compact sources with those in the TGSS-ADR1 150 MHz survey \citep{Intema2017}. The selected sources have a flux densities above $20\sigma$ in both LOFAR and TGSS-ADR1 images. We found that the flux scale in our LOFAR image is about 4 percent higher than the scale in the TGSS-ADR1 image. We adapt in our analysis a flux scale uncertainty of 10~percent.

\begin{table}
	\caption{Imaging parameters.}
	\begin{tabular}{lcccc}
		\hline\hline
		Data          &   $uv$-range   &   $\mathtt{Robust}^a$   &   $\theta_{\rm \tiny FWHM}$   &     $\sigma$         \\
		& (k$\lambda$) & ($\mathtt{uv-taper}$) & ($\arcsec\times\arcsec$, $PA^b$) & ($\upmu{\rm Jy}\,{\rm beam}^{-1}$)                                \\ \hline
		LOFAR &   $0.15-65$   &  $-0.25$   &   $16.6\times8.9$ &                80\\
		&&($5\arcsec$) & ($84^\circ$)   & \\
		&   $0.15-65$   &  $0.25$ &         $60.4\times46.7$         &                180\\ 
		&& ($10\arcsec$) & ($-73^\circ$)& \\
		VLA   &   $0.52-74.6$   &  $0.25$  &         $10.9\times8.6$      &                 38 \\ 
		&&($10\arcsec$)  &($68^\circ$)    &           \\ 
		\hline\hline
	\end{tabular}\\	
	Notes: $^a$: Briggs weighting of the visibility. $^b$: position angle.
	\label{tab:image_para}
\end{table}

\subsection{VLA data}
\label{sec:jvla} 

The VLA L band ($1-2\,{\rm GHz}$) data was taken on February 22, 2015 while the telescope was in the B-array configuration. The target was observed for four scans of approximately ten minutes each, with a few hours in between the scans.  The calibration is processed in the standard fashion (flux, phase, amplitude, band-pass etc.) with $\mathtt{CASA}$\footnote{\url{https://casa.nrao.edu}} using 3C~147 and 3C~286 as calibrators. The flux density scale was set according to the \cite{Perley2017} flux scale.
The RFI was automatically removed by the $\mathtt{CASA}$ `tfcrop' and `flagdata' tasks. To eliminate the remaining RFI after the $\mathtt{CASA}$ flaggers, the $\mathtt{aoflagger}$ \citep{Offringa2012c} was employed. Additionally, outlier visibilities were looked for in the amplitude-$uv$ distance plane by inspecting the cross-hand (RL, LR) correlations. Since most radio sources are generally not polarised, the cross-hand correlations are a good additional indicator of RFI. In this plane, visibilities more than $6\sigma$ away from the mean were flagged. 

Finally, to remove residual amplitude and phase errors and increase the quality of the image, we performed self-calibration of the target field. Three rounds of phase calibration and three rounds of amplitude and phase calibration with the decrease of the solution interval each round were done. Amplitude solutions with a signal-to-noise ratio smaller than three were flagged, as well as outlier ($>3\sigma$ away from mean) amplitude solutions. The remaining RFI is manually flagged before the calibrated data is imaged with $\mathtt{WSClean}$ \citep{Offringa2014}. The final images are corrected for the attenuation of the primary beam by dividing the image pixel values by the VLA primary beam model created by $\mathtt{CASA}$.

\subsection{Chandra data}
\label{sec:chandra} 

Chandra observations of A990 were performed  for 10~ks on June 03, 2013 with the AXAF CCD Imaging Spectrometer (ObsID: 15114). The data were calibrated using the $\mathtt{ClusterPyXT}$ automatic pipeline\footnote{\url{https://github.com/bcalden/ClusterPyXT}} \citep{Alden2019} that calls pre-built routines in $\mathtt{CIAO}$\footnote{\url{https://cxc.harvard.edu/ciao}}\citep{Fruscione2006} and $\mathtt{Sherpa}$\footnote{\url{https://cxc.cfa.harvard.edu/sherpa/}}\citep{Freeman2001}. The pipeline is able to generate X-ray surface brightness, spectral temperature, pressure, particle density, and entropy maps from Chandra observations with minimal inputs from users. The pipeline retrieves the data and corresponding backgrounds from the Chandra data archive. As our interest is X-ray extended emission from the cluster, point sources are manually removed from the data. High energy events are identified and filtered out by the pipeline. The gas temperature and pressure are obtained by spectral fitting of the events in the regions of interest. $\mathtt{ClusterPyXT}$ uses  Astronomical Plasma Emission Code ($\mathtt{APEC}$) model for optically thin collisionally ionized host plasma and a photoelectric absorption model (i.e. $\mathtt{X{\rm -}Spec~PHABS}$) that includes redshift, metallicity, temperature, normalization and hydrogen column density along the line of sight \citep{Balucinska-Church1992,Smith2001}. We use a metallicity of $0.3\,Z_\odot$ \citep[e.g.][]{Werner2013} and a hydrogen column density of $1.31\times10^{20}\,{\rm cm}^{-2}$ \citep{Bekhti2016} for the ICM. For details of the reduction procedure, we refer to \cite{Alden2019}.

\section{Results}

\subsection{Radio emission}
\label{sec:radio}

\begin{figure*}
	\centering
	\includegraphics[width=1.9\columnwidth]{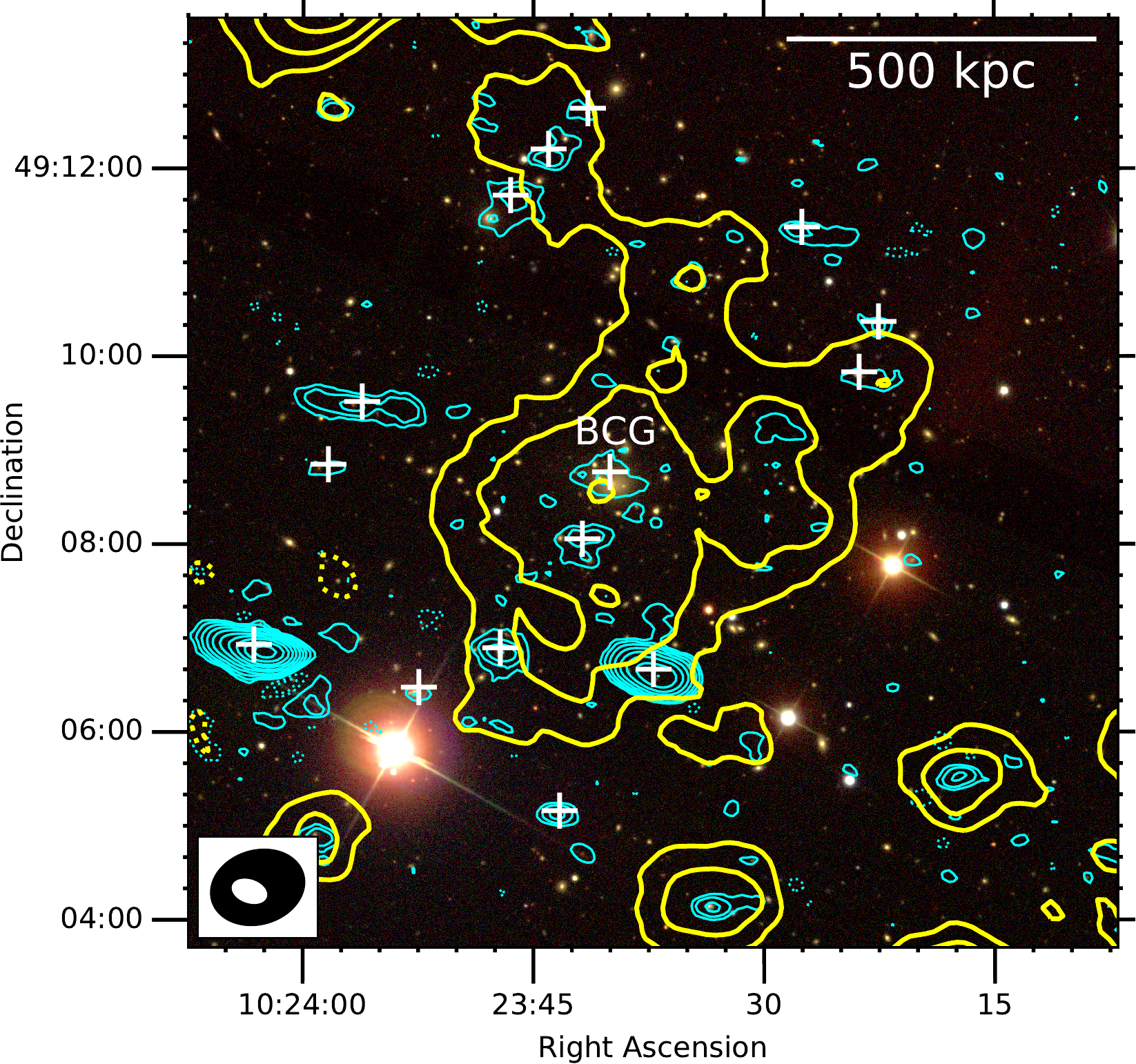}
	\caption{LOFAR 144 MHz high- ($16.6\arcsec\times8.9\arcsec$, cyan) and low- ($60.4\arcsec\times46.7\arcsec$, yellow) resolution contours  on the SDSS optical (i, r, and g band) image. The contours are drawn at levels of $\pm[1,2,4,8,16,32] \times3\sigma$, where $\sigma=80\,\upmu{\rm Jy\,beam}^{-1}$ and $\sigma=180\,\upmu{\rm Jy\,beam}^{-1}$ for the high- and low-resolution contours, respectively. Negative contours are shown by the dotted lines. Compact sources including the BCG marked with the plus (+) are subtracted in the low-resolution contours. The synthesis beams are shown in the bottom left corner. Source labels are given in Figure \ref{fig:a990_label}.}
	\label{fig:a990_lofar}
\end{figure*}

\begin{figure}
	\centering
	\includegraphics[width=1\columnwidth]{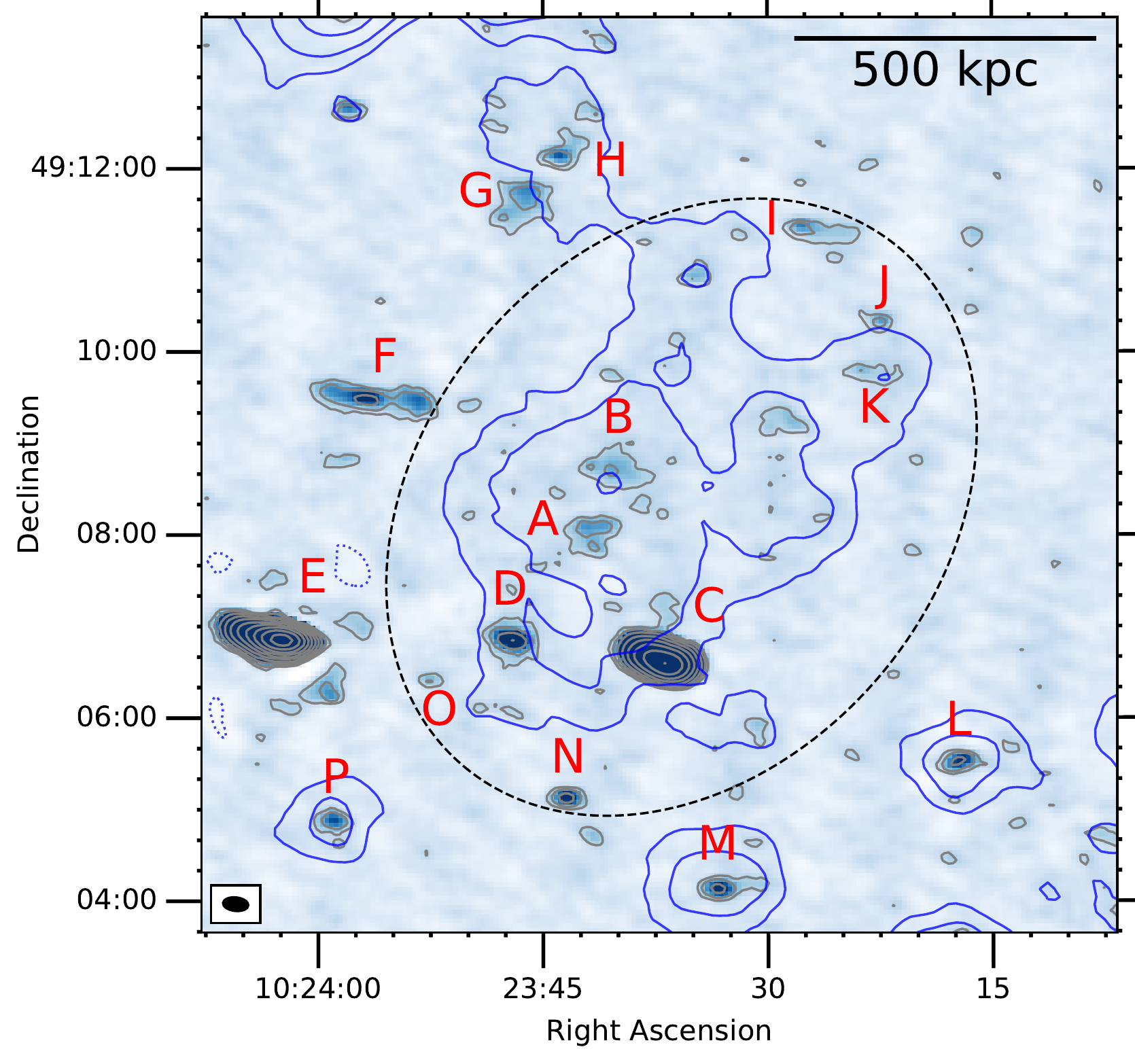}
	\caption{LOFAR 144 MHz high-resolution image of A990. The high-(gray) and low-(blue) resolution contours are the same as those in Figure \ref{fig:a990_lofar}. The ellipse is the region where the flux density of the halo is integrated. Compact sources are labelled.}
	\label{fig:a990_label}
\end{figure}

In Figure~\ref{fig:a990_lofar} we present the LOFAR 144 MHz radio emission contour maps of A990 at high and low resolution (see Table~\ref{tab:image_para} for the image properties). The high-resolution contours show the detection of radio emission from galaxies which are labelled in Figure \ref{fig:a990_label}. The radio emission from the brightest cluster galaxy \citep[BCG;  at $\alpha=10^{\rm h}23^{\rm m}39.9^{\rm s}$, $\delta=+49^{\rm d}08^{\rm m}38.4^{\rm s}$ and redshift $z_{sp}=0.14216\pm0.00005$;][]{Alam2015} is also seen in the high-resolution image. Due to the low surface brightness of the cluster-scale emission, it is only seen in the low-resolution contours. In the subsections below, we present measurements for the detected sources.

\subsubsection{Extended radio source}
\label{sec:flux_spec}

Figure 1 shows that the extended radio emission is detected in the cluster centre, and its morphology is mainly extended in the NW$-$SE direction. The minimum and maximum projected sizes within the $3\sigma$ contour are 515~kpc and 960~Mpc, respectively. The projected size of the extended source is  approximated as $D=\sqrt{515\,{\rm kpc}\times960\,{\rm kpc}}\approx700\,{\rm kpc}$.  The location and size of the extended source that mostly overlays the X-ray extended emission suggest that it is a radio halo.

The integrated flux density for the radio halo measured in the LOFAR low-resolution map is $S_{\rm 144~MHz}=20.2\pm2.2 \,{\rm mJy}$. Here we select only pixels that are detected with $\geq3\sigma$ in the elliptical region  in Figure \ref{fig:a990_label}. We did not integrate the emission in the region of source G and H as this is likely the residuals of the imperfect subtraction of these sources. The error  takes into account the image noise and the flux scale uncertainty of 10~percent which are added in quadrature. The corresponding 144~MHz power for the radio halo at $z=0.144$ is $P_{\rm 144~MHz}=(1.2\pm0.1)\times10^{24}~{\rm W~Hz}^{-1}$ ($k-$corrected, assuming a spectral index of $\alpha=-1.3$, i.e. the mean indices for a number of known halos, \citealt{Luigina2012}). To the best of our knowledge, A990 hosts one of the lowest radio power halos at low ($\lesssim300$~MHz) frequencies known to date \citep[together with RXC~J1825.3+3026 studied by ][]{Botteon2019}.  If the halo has a steep spectrum, $\alpha\lesssim-1.3$, the radio power at 1.4~GHz is $P_{\rm 1.4~GHz}\lesssim(6.1\pm0.7)\times10^{22}~{\rm W~Hz}^{-1}$ which is still being one of the lowest powers for a radio halo at 1.4~GHz \citep[][]{Botteon2019,Birzan2019}. When all pixels above $2\sigma$ are included, the values  increase by about 12 percent to $S_{\rm 144~MHz}=22.7\pm2.4 \,{\rm mJy} $ and $P_{\rm 144~MHz}=(1.3\pm0.1)\times10^{24}~{\rm W~Hz}^{-1}$. A similar difference in the flux density measurements within $2\sigma$ and $3\sigma$ contours was also seen in PSZ2~G099.86+58.45 \citep[i.e. 15 percent;][]{Cassano2019}. In this paper, we adopt the flux density measured with the  $\geq3\sigma$ pixels for the radio halo in A990 (i.e. $S_{\rm 144~MHz}=20.2\pm2.2 \,{\rm mJy}$).

\begin{figure*}
	\centering
	\includegraphics[width=0.33\textwidth]{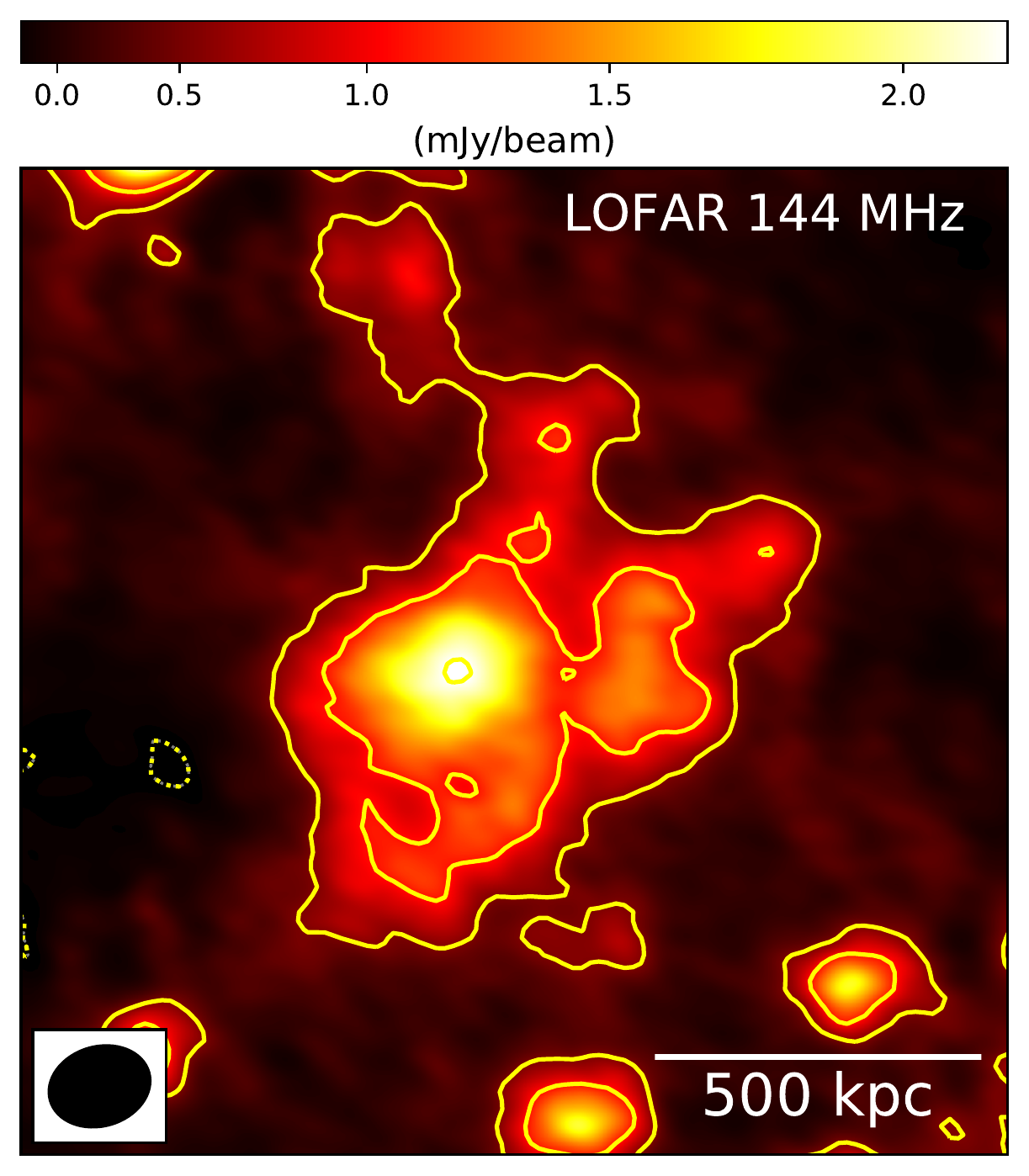} \hfill
	\includegraphics[width=0.33\textwidth]{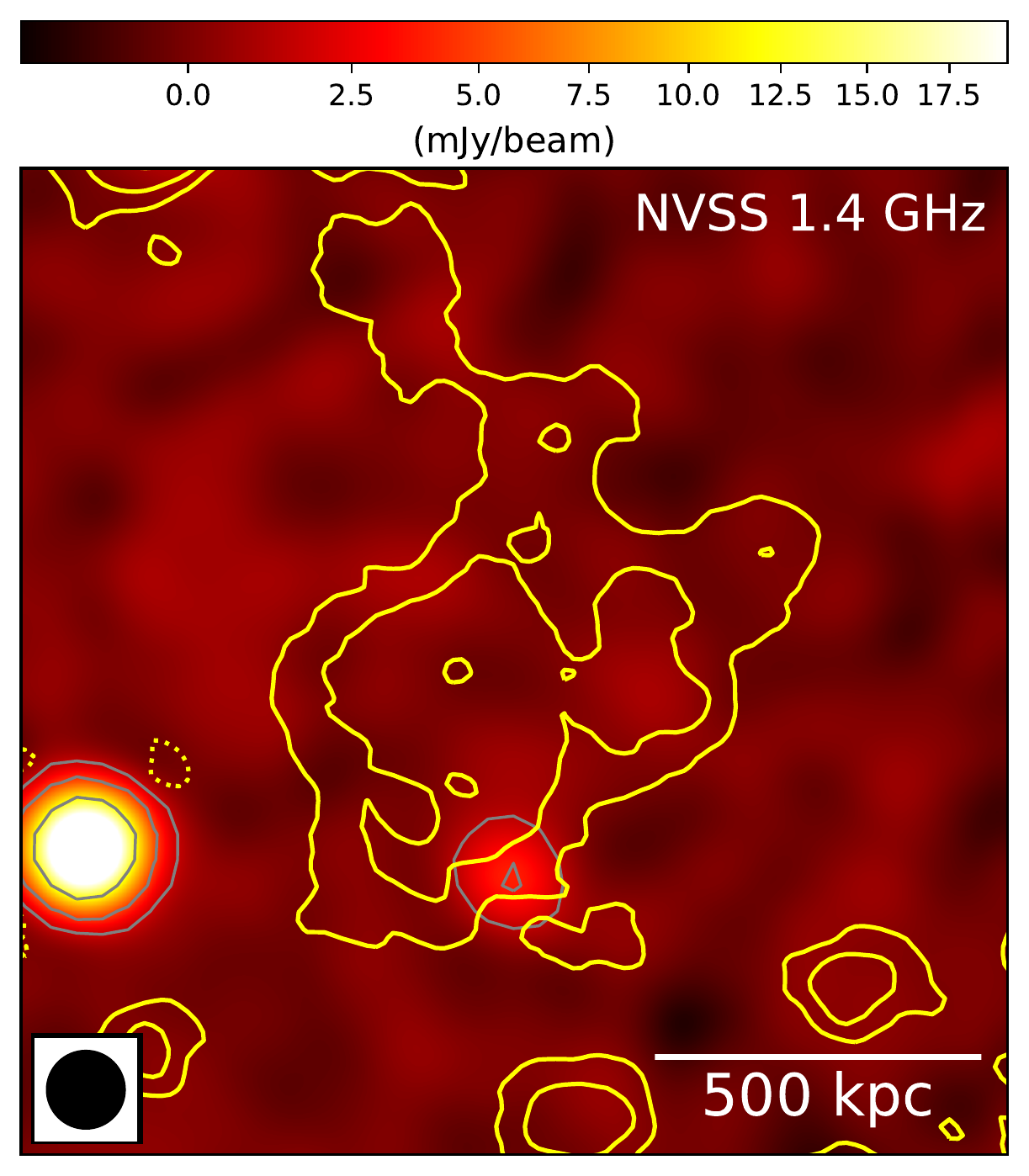} \hfill
	\includegraphics[width=0.33\textwidth]{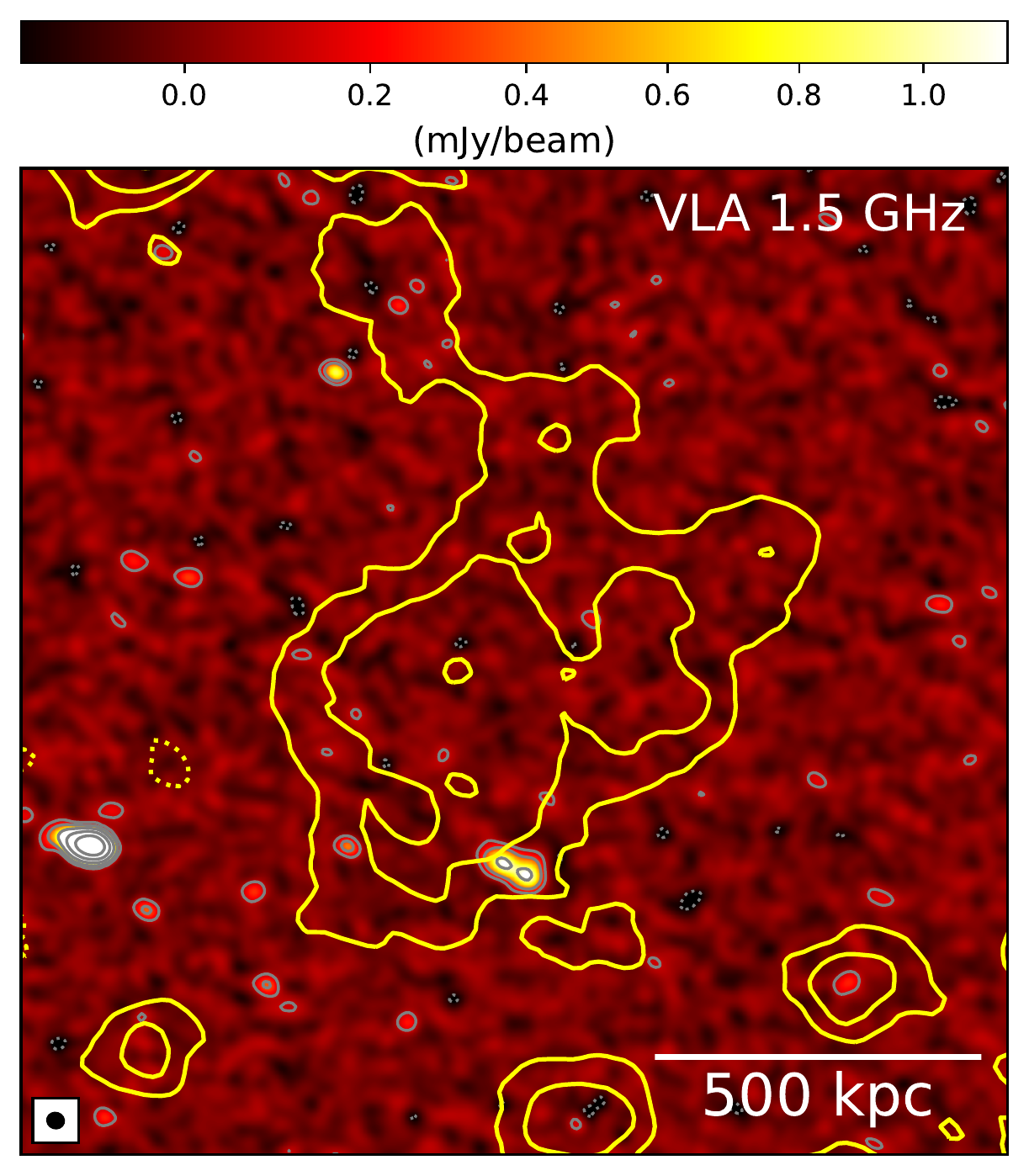}
	\caption{ (From left to right) LOFAR 144~MHz, NVSS 1.4~GHz, and VLA 1.5~GHz image of A990. The NVSS and VLA (grey) contours are $\pm[1,3,9,27,81] \times3\sigma$, where $\sigma=450\,\upmu{\rm Jy\,beam}^{-1}$ and $38\,\upmu{\rm Jy\,beam}^{-1}$ for the NVSS and VLA images, respectively. The field of view of the images and the LOFAR low-resolution (yellow) contours without compact sources are the same as those in Figure \ref{fig:a990_lofar}.}
	\label{fig:a990_lnv}
\end{figure*}

The radio halo is undetected in the NVSS and VLA high-frequency  images in Figure~\ref{fig:a990_lnv}. The non-detection could be due to the sensitivity of the NVSS and VLA data. To examine this possibility, we compute the expected surface brightness at $1.4\,{\rm GHz}$ and $1.5\,{\rm GHz}$, and compare it with the  measurements in the NVSS and VLA images. Using the flux density measurement from the LOFAR data, we estimate the flux densities for the halo to be $S_{\rm 1.4~GHz}=2.1\,{\rm mJy}$ and $S_{\rm 1.5~GHz}=2.0\,{\rm mJy}$. Here we assume a flat spectrum, $\alpha=-1.0$, for the halo. This is a conservative approach as radio halos are generally found to have steeper spectra  \citep[$\bar{\alpha}\approx-1.3$;][]{Luigina2012}. The true flux densities of the halo at 1.4~GHz and 1.5~GHz could be lower than these estimates (e.g. with $\alpha=-1.3$,  $S_{\rm 1.4~GHz}=1.1\,{\rm mJy}$ and $S_{\rm 1.5~GHz}=1.0\,{\rm mJy}$). To calculate the expected surface brightness of the halo, we assume that the halo emission is uniformly distributed over the source area. Hence, the surface brightness is $I_{\rm 1.4~GHz}=\nicefrac{S_{\rm 1.4~GHz}}{A}=0.035\,\upmu{\rm Jy\,arcsec}^{-2}$ and $I_{\rm 1.5~GHz}=0.033\,\upmu{\rm Jy\,arcsec}^{-2}$, where $A=61,075\,{\rm arcsec}^2$ is the area of the halo (i.e. covered by the $\geq3\sigma$ in the LOFAR map). However, the sensitivity for the  NVSS and VLA images is $\sigma=0.20\,\upmu{\rm Jy\,arcsec}^{-2}$ \citep[][]{Condon1998} and $\sigma=0.36\,\upmu{\rm Jy\,arcsec}^{-2}$ (see Table \ref{tab:image_para}), respectively. These are not sensitive enough to detect the halo in A990. However, the radio emission at 144~MHz is not constant over the halo (Figure \ref{fig:a990_lnv}). The surface brightness emission at the peak region is $\sim$$1.7$ times the mean pixel value in the halo region. In case that the emission at 1.4~GHz and 1.5~GHz follows the similar structure at 144~MHz, the surface brightness in some regions of the halo is expected to be higher, i.e. up to $0.06\,\upmu{\rm Jy\,arcsec}^{-2}$  and $0.05\,\upmu{\rm Jy\,arcsec}^{-2}$, respectively. The expected surface brightness is still below the detection limit of the NVSS and VLA observations. 
We also note that the NVSS and VLA observations are not ideal for detecting faint, extended emission at the scales of the cluster. The NVSS data obtained from snapshot observations is poorly sampled in the $uv$ space. The VLA~1.5~GHz observations in B configuration is also lacking of short baselines.
To estimate the halo spectrum, future deep observations at high frequencies with the VLA (e.g. C, D configuration) and uGMRT or at low frequencies with LOFAR Low Band Antennas (10--80~MHz) will be necessary.

\subsubsection{Compact radio sources}
\label{sec:compact}

\begin{table*}
	\caption{Flux densities and spectra for radio compact sources.}
	\begin{tabular}{lcccc}
		\hline\hline
		Source & $S_{\rm 144\,MHz}$ [mJy] & $S_{\rm 1.4\,GHz}$ [mJy] &    $\alpha$    &           $z$           \\ \hline
		A      &      $1.77\pm0.19$       &           $0.04\pm0.04$            &       $-1.66\pm0.41$      &           --            \\
		B      &      $1.79\pm0.20$       &            --            &       --       &      $0.14216\pm0.00005^{a}$      \\
		C      &      $75.10\pm7.51$      &      $4.27\pm0.22$       & $-1.26\pm0.05$ &  $0.6283\pm0.0356^{a}$  \\
		D      &      $3.07\pm0.32$       &      $0.34\pm0.04$       & $-0.97\pm0.07$ &  $0.1326\pm0.027^{a}$   \\
		E      &     $255.57\pm25.46$     &      $34.74\pm1.74$      & $-0.88\pm0.05$ &           --            \\
		F      &      $3.08\pm0.32$       &      $0.44\pm0.04$       & $-0.86\pm0.06$ &      $0.5383\pm0.0303^{a}$       \\
		G      &      $2.33\pm0.25$       &      $0.71\pm0.05$   &    $-0.53\pm0.06$ &  -- \\ 
		H      &      $1.26\pm0.15$       &      $0.18\pm0.04$       & $-0.87\pm0.11$ &  -- \\ 
		I      &      $1.31\pm0.15$       &      $0.01\pm0.04$       & $-2.18\pm1.81$ &           --            \\
		J      &      $0.44\pm0.09$       &            --            &       --       &  $0.6717\pm0.0655^{a}$  \\
		K      &      $0.59\pm0.10$       &            --            &       --       &  $0.1326\pm0.0364^{a}$  \\
		L      &      $1.24\pm0.15$       &      $0.24\pm0.04$       &$-0.73\pm0.09$ &           --            \\
		M      &      $1.52\pm0.17$       &      $0.02\pm0.04$       & $-1.88\pm0.79$ &  $0.1219\pm0.0206^{a}$  \\
		N      &      $1.08\pm0.13$       &      $0.14\pm0.04$       & $-0.92\pm0.14$ &           --            \\
		O      &      $0.24\pm0.08$       &      $0.21\pm0.04$      & $-0.06\pm0.17$ &  $0.095\pm0.0481^{a}$   \\
		P      &      $0.93\pm0.12$      &            --            &       --       & $0.15073\pm0.00012^{b}$ \\ \hline\hline
	\end{tabular}\\	
	References: $^a$: \cite{Alam2015}; $^b$: \cite{Rines2013}.
	\label{tab:ps}
\end{table*}

A number of compact radio sources are detected with our LOFAR observations in Figure \ref{fig:a990_lofar}.  In Table~\ref{tab:ps} we give the flux densities and spectra between 144~MHz and 1.5~GHz for the sources within a field of view of $\sim$$10\arcmin$ centred on the cluster. The spectral indices of the compact sources range from $-0.50$ to $-1.84$. The redshifts of sources B, D, K, and P range from $0.1326$ to $0.15073$ implying that they are at similar redshifts of the cluster members ($\bar{z}=0.144$). The overlaid radio-SDSS optical images of the compact sources are shown in Figure \ref{fig:a990_cutouts}.  Sources A , E, G, H, I, L, and N do not have clear SDSS optical counterparts. Sources A and B consist of multiple point sources. Sources G and H are likely two lobes of a FR-I type active galactic nucleus. 

\subsection{X-ray emission}
\label{sec:xray}

\begin{figure}
	\centering
	\includegraphics[width=1\columnwidth]{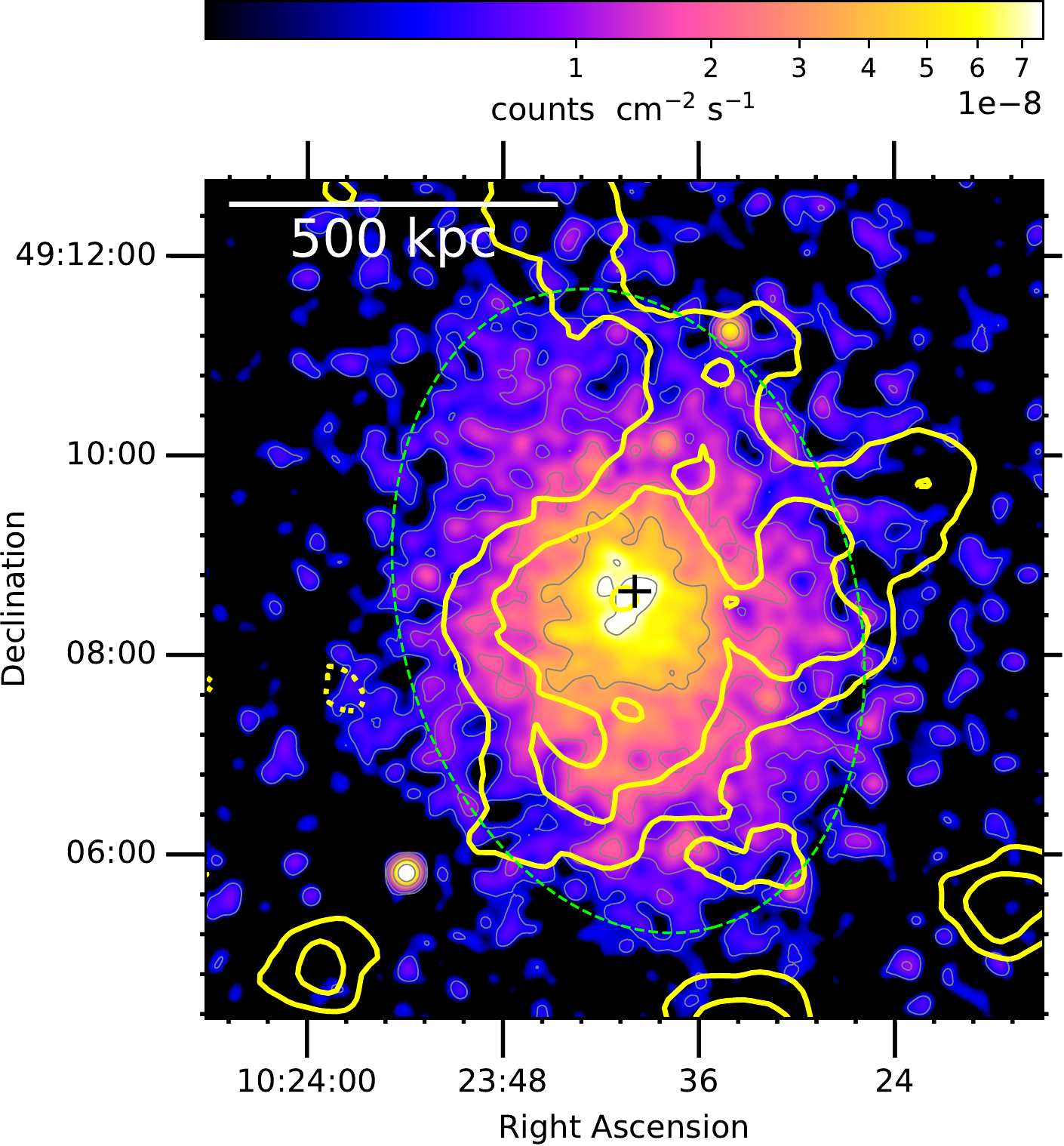}
	\caption{\textbf{Chandra $0.5-2.0$~keV image} (bin=2) of the cluster overlaid with the LOFAR low-resolution contours. The Chandra data is smoothed with a Gaussian kernel of 5 pixels where the pixel size is $0.98\arcsec$. The X-ray emission peaks at the cross (+) location (${\rm RA}=10^{\rm h}23^{\rm m}39.9^{\rm s}$ and ${\rm Dec}=+49^{\rm d}08^{\rm m}38.4^{\rm s}$). The ellipse used for X-ray luminosity measurement is centred at ${\rm RA}=10^{\rm h}23^{\rm m}40.4^{\rm s}$, ${\rm Dec}=+49^{\rm d}08^{\rm m}25.7^{\rm s}$ with a position angle of 15 degree.  The major and minor axes are $4.58\arcmin$ and $6.57\arcmin$ (i.e. $694\,{\rm kpc}$  and $996\,{\rm kpc}$), respectively.}
	\label{fig:a990_chandra}
\end{figure}

The Chandra $0.5-2.0\,{\rm keV}$ image of the cluster is shown in Figure~\ref{fig:a990_chandra}.  The X-ray emission from the cluster is relatively concentrated in the central region. Its peak emission is closed to the location of the BCG. In the outskirts, the X-ray emission has a lower surface brightness and is elongated in the NE$-$SW direction, which is  perpendicular to the major axis of the LOFAR 144~MHz radio emission. In the NE peripheral region of the cluster, X-ray emission is found, but radio emission is not detected. Reversibly, extended radio emission is detected in the west side of the cluster without X-ray emission seen. The total X-ray luminosity between energy range 0.1~keV and 2.4~keV within an elliptical region in Figure \ref{fig:a990_chandra} is $L_{\rm X} = (3.66\pm0.08)\times10^{44}\,{\rm ergs\,s}^{-1}$, implying that A990 is one of the lowest X-ray luminosity clusters ever found to host a radio halo \citep[][]{Birzan2019}. In the energy range 0.5--2.0~eV, the X-ray luminosity is $L_{\rm X} = (2.45\pm0.06)\times10^{44}\,{\rm ergs\,s}^{-1}$. The elliptical region is chosen to roughly cover the X-ray $3\sigma$ contours (i.e. $\sigma=1.5\times10^{-9}\,{\rm counts\,cm^{-2}s^{-1}}$  ). At the measured X-ray luminosity the cluster is about 8 times underluminous in the radio power compared with the $L_{\rm X}-P_{\rm 1.4~GHz}$ correlation \citep[e.g.][]{Cassano2013}.

To study dynamical status of the cluster,  we use the X-ray data to derive  morphological indicators and to construct a cluster temperature map.

\begin{figure}
	\centering
	\includegraphics[width=1\columnwidth]{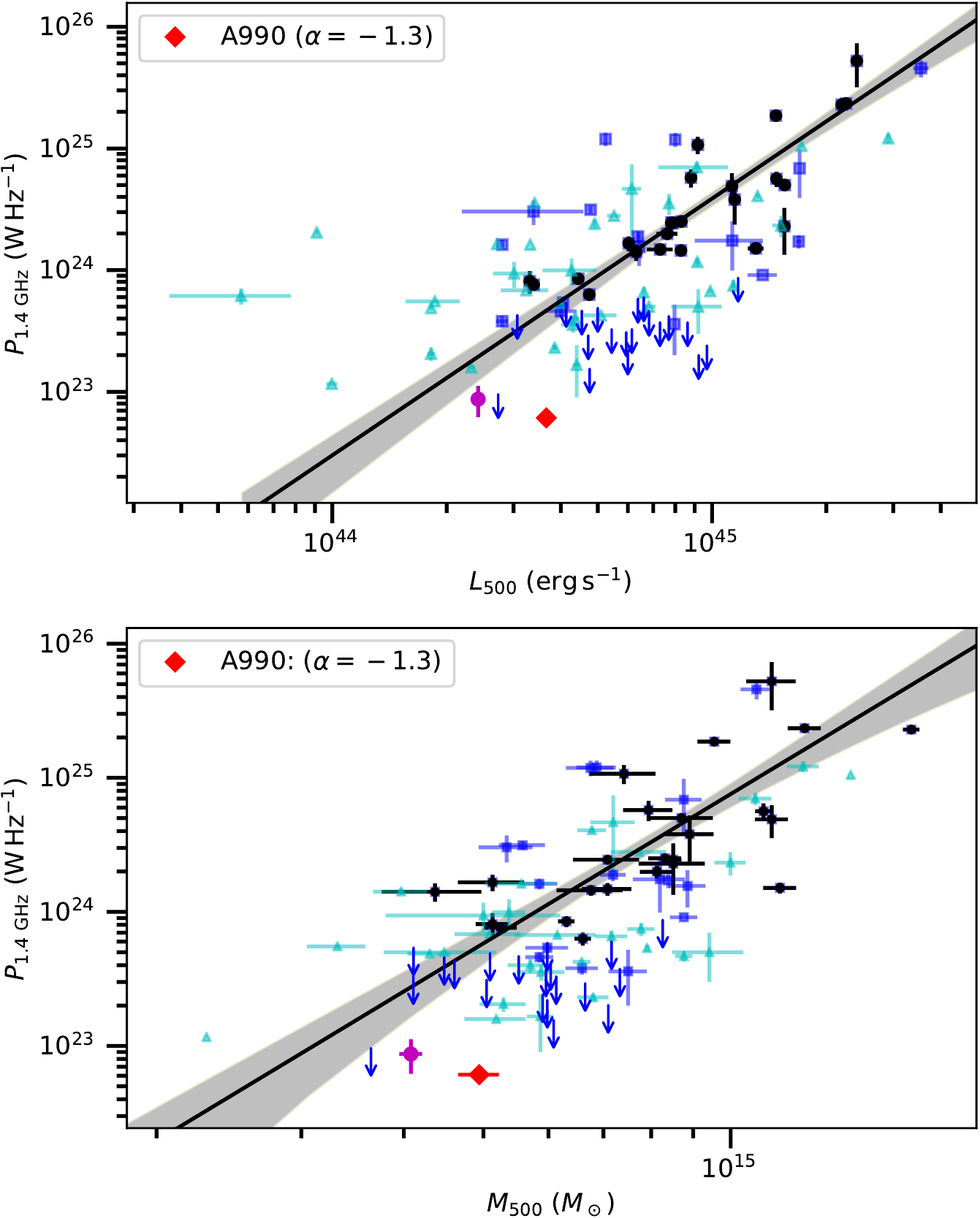}
	\caption{The $P_{\rm 1.4~GHz} - M_{\rm 500}$ (bottom) and $P_{\rm 1.4~GHz}-L_{500}$ (top) relations, adapted from \protect\cite{Birzan2019}. The data points for A990 and RXC~J1825.3+3026  are shown by the red diamond and magenta circle, respectively. In both panels, the black circles, blue squares, and cyan triangles show radio halos in the samples of \protect\cite{Cassano2013}, \protect\cite{MartinezAviles2016}, and \protect\cite{Birzan2019}, respectively. The blue arrows indicate the power upper limits for undetected halos in the \protect\cite{Cassano2013}, \protect\cite{Kale2015}, and  \protect\cite{Birzan2019} cluster samples. The solid lines are the best fits for the \protect\cite{Cassano2013} halos. The shadowed regions show the 95 percent confidence region.}
	\label{fig:correlations}
\end{figure}

\subsubsection{Dynamical status}
\label{sec:dynamics}

Morphological properties of X-ray emission are known to correlate with dynamical state of the host clusters of galaxies \citep[e.g.][]{Cassano2010,Bonafede2017}. The X-ray emission from A990 in Figure~\ref{fig:a990_chandra} is mostly concentrated in the central region and is slightly elongated in the NE$-$SW direction which seems to indicate a slightly disturbed cluster but without major merging activities. We estimate a set of  parameters  to classify the merging status. These include (i) the concentration of the X-ray emission $c$ \citep{Santos2008},  (\textit{ii}) the centroid shifts between the X-ray emission peak and the centroid of the X-ray emission within aperture sizes \citep[e.g.][]{Poole2006}, and (\textit{iii}) the power ratios $\nicefrac{P_m}{P_0}$ \citep[$m=3$;][]{Buote1995}. These are briefly described below.

\begin{description}
	\item[\textit{i}.] Concentration of X-ray emission is defined as $c=\nicefrac{S_{\rm X}(r_1<100~{\rm kpc})} {S_{\rm X}(r_2<500~{\rm kpc})}$, where $S_{\rm X}(r_i)$ is the total X-ray surface brightness within radius $r_i$.
	
	\item[\textit{ii}.] Centroid shift $w=\nicefrac{SD[\Delta_i(r_i)]}{R_{{\rm ap}}}$,  where $\Delta_{i}$ is the projected separation between the X-ray emission peak and the  centroid of the \textit{i}$^ {\rm th}$ aperture of radius $r_i$. 
	
	\item[\textit{iii}.] Power ratio $\nicefrac{P_m}{P_0}$ presents the fraction of the \textit{m}$^{\rm th}$ multipole moment to the total power gravitational potential. The lowest power ratio moment with $m=3$ gives a measure of the cluster substructures \cite[e.g.][]{Bohringer2010}. For detailed formula, we refer to \cite{Buote1995}.
	
\end{description}

Several studies have used these morphological parameters to constrain the dynamical state of galaxy clusters \citep[e.g.][]{Bohringer2010,Cassano2010,Bonafede2017,Savini2019}. Clusters that are dynamically disturbed and host radio halos have a low value of $c$, high values of $w$ and $\nicefrac{P_3}{P_0}$ \citep[e.g.][]{Cassano2010,Cuciti2015}, except some cases in \cite{Bonafede2014c,Bonafede2015}. 

\begin{figure*}
	\centering
	\includegraphics[width=0.32\textwidth]{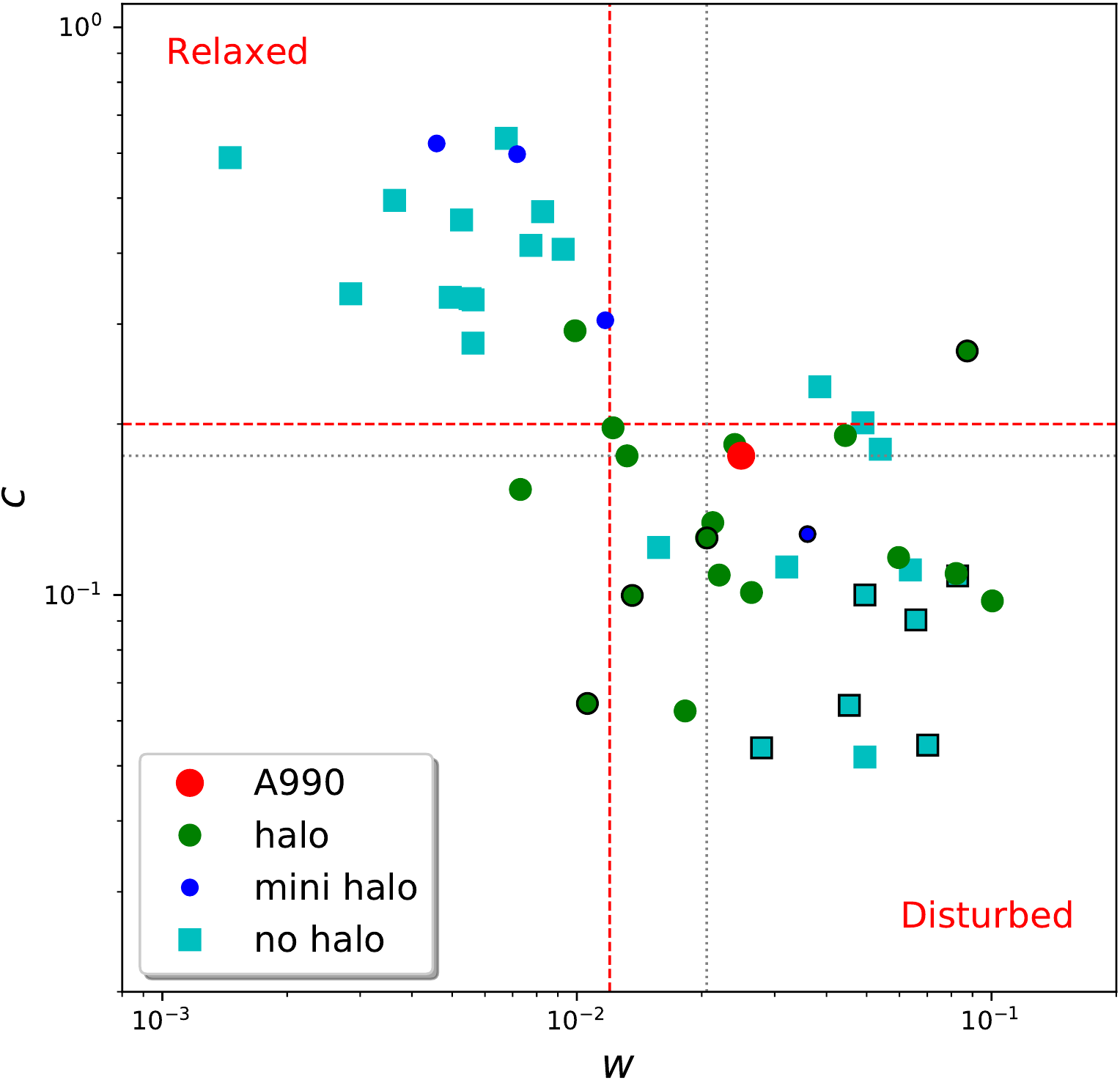} \hfill
	\includegraphics[width=0.32\textwidth]{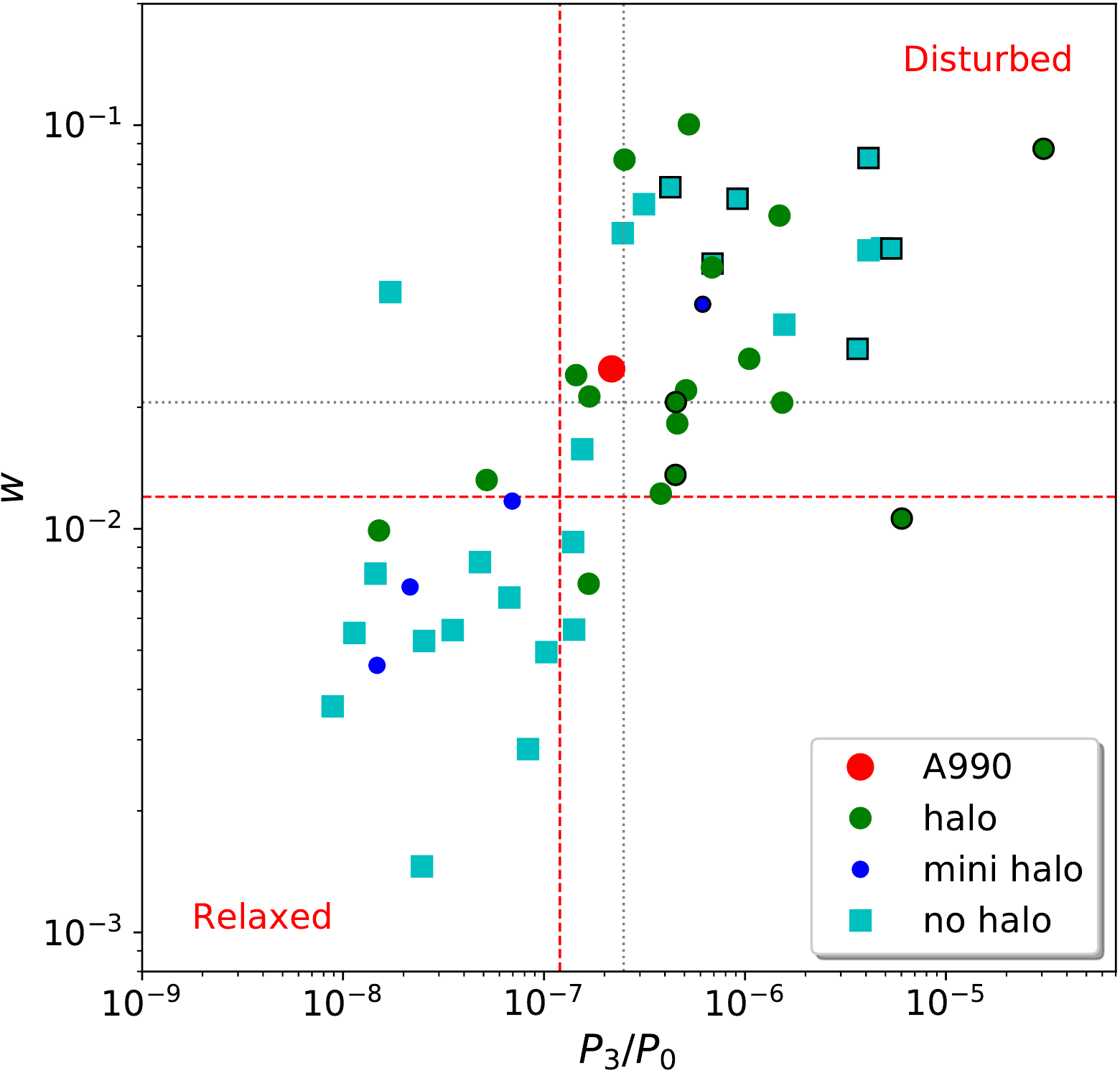} \hfill
	\includegraphics[width=0.32\textwidth]{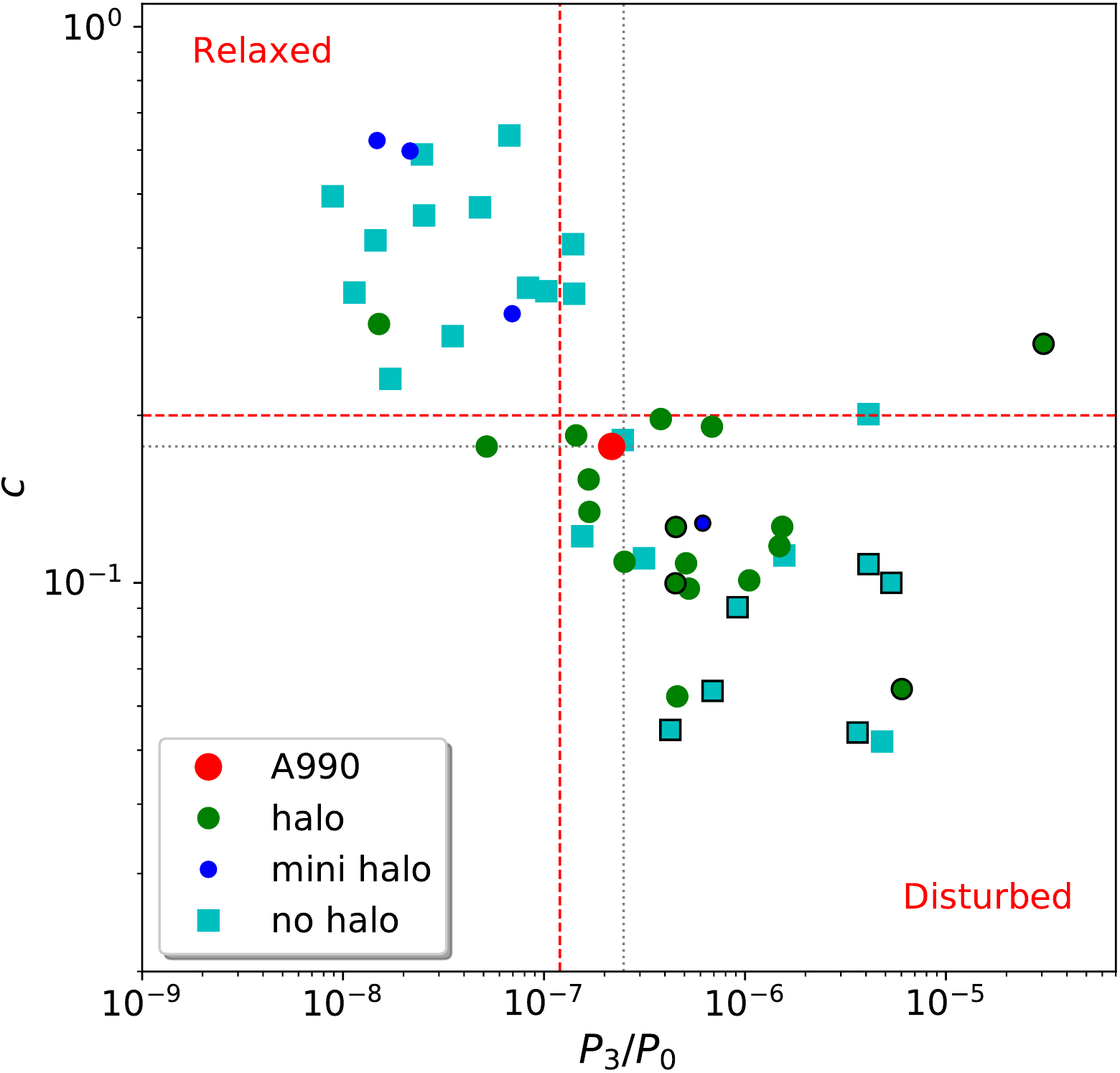} 
	\caption{Morphological parameter diagrams for clusters with and without radio halos \protect\citep{Cassano2010,Bonafede2014c,Bonafede2017}. Clusters with radio halos are shown with filled circles. Radio quiet clusters are plotted with cyan squares. Clusters hosting double relics are additionally marked with black circles. The data points for A990 are  marked with the red circles. The red dashed lines indicate the median values (i.e. $c=0.2$, $w=0.012$, $\nicefrac{P_3}{P_0}=1.2\times10^{-7}$) calculated from a complete sample of 32 galaxy clusters \protect\citep{Cassano2010}. The black dotted lines (i.e. $c=0.18$, $w=0.021$, $\nicefrac{P_3}{P_0}=2.5\times10^{-7}$) are calculated for all clusters in the plots.}
	\label{fig:a990_indicators}
\end{figure*}

Following the calculations in \cite{Cassano2010,Bonafede2017}, we estimate the morphological parameters for A990 to be $c=0.18$, $w=0.025$, and $\nicefrac{P_3}{P_0}=2.17\times10^{-7}$. The X-ray emission peak used in the calculation is shown in Figure \ref{fig:a990_chandra}.  The centroid shift is calculated for circular apertures within $R_{{\rm ap}}=500~{\rm kpc}$. The first aperture  starts from $r_i=0.05R_{{\rm ap}}$ and the next ones increase in steps of $0.05R_{{\rm ap}}$. The power ratio $\nicefrac{P_3}{P_0}$ is  also calculated out to a radius of $R_{{\rm ap}}$. We add our results to the morphological parameter diagrams in Figure~\ref{fig:a990_indicators}. 

The morphological parameter diagrams in Figure~\ref{fig:a990_indicators} show that A990 is located in the morphologically disturbed quadrant separated by the medians of $c$, $w$, and $\nicefrac{P_3}{P_0}$ calculated from a sample of 32 galaxy clusters with X-ray luminosity $L_{X}\geq5\times10^{44}\,{\rm erg\,s^{-1}}$ and redshift $0.2\leq z \leq0.32$ \citep{Cassano2010}. When adding more clusters from \cite{Bonafede2015a} and from double relics \cite{Bonafede2017}, the medians are slightly shifted (see Figure \ref{fig:a990_indicators}) and the points for A990 are closed to the boundaries for disturbed and relaxed clusters. This is in line with the idea that radio halos are generated in disturbed galaxy clusters that supply energy via turbulence for the particle re-acceleration and magnetic field amplification/generation.

\subsubsection{Temperature map}
\label{sec:temp}

\begin{figure*}
	\centering
	\includegraphics[width=0.49\textwidth]{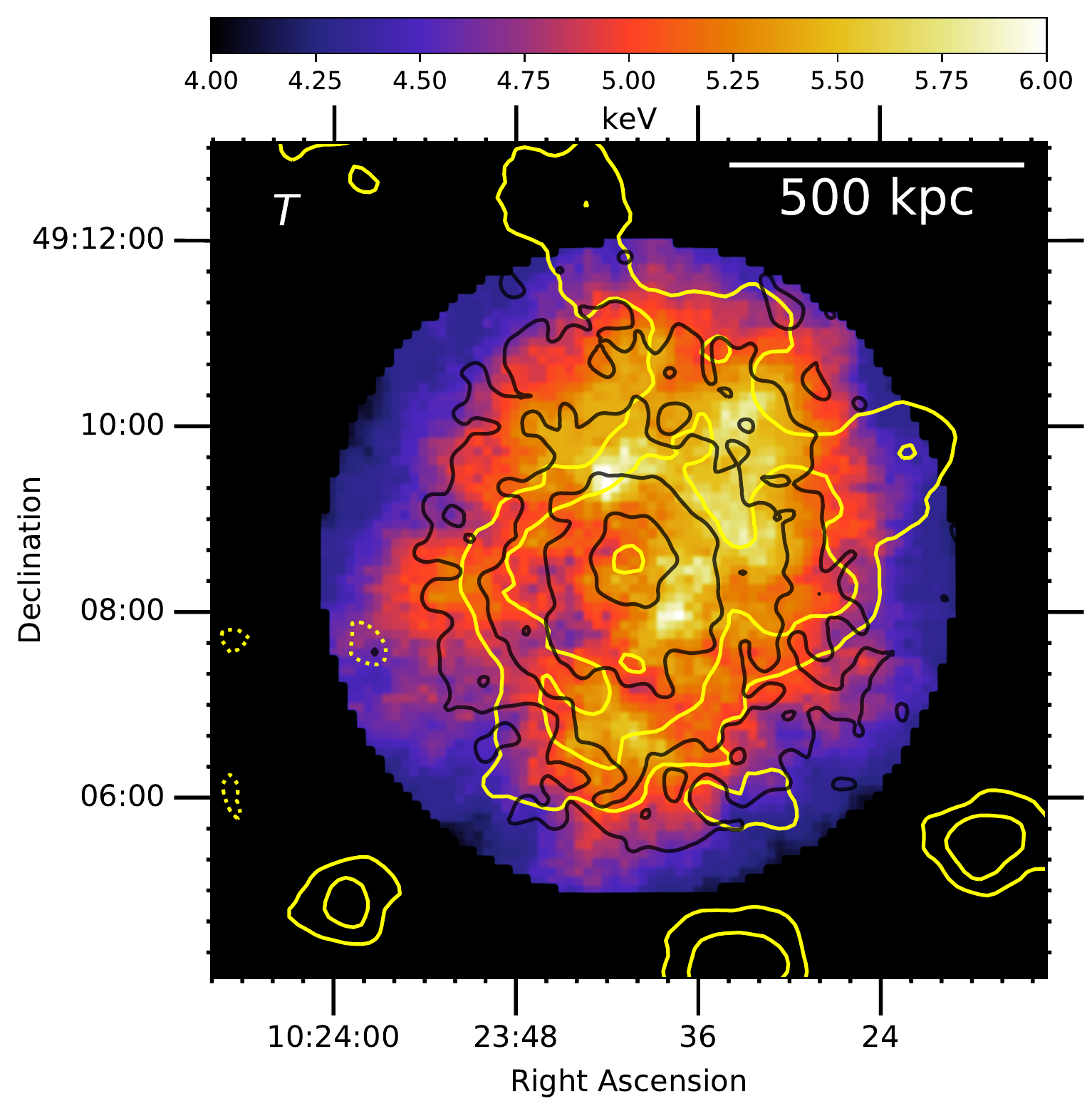} \hfill
	\includegraphics[width=0.49\textwidth]{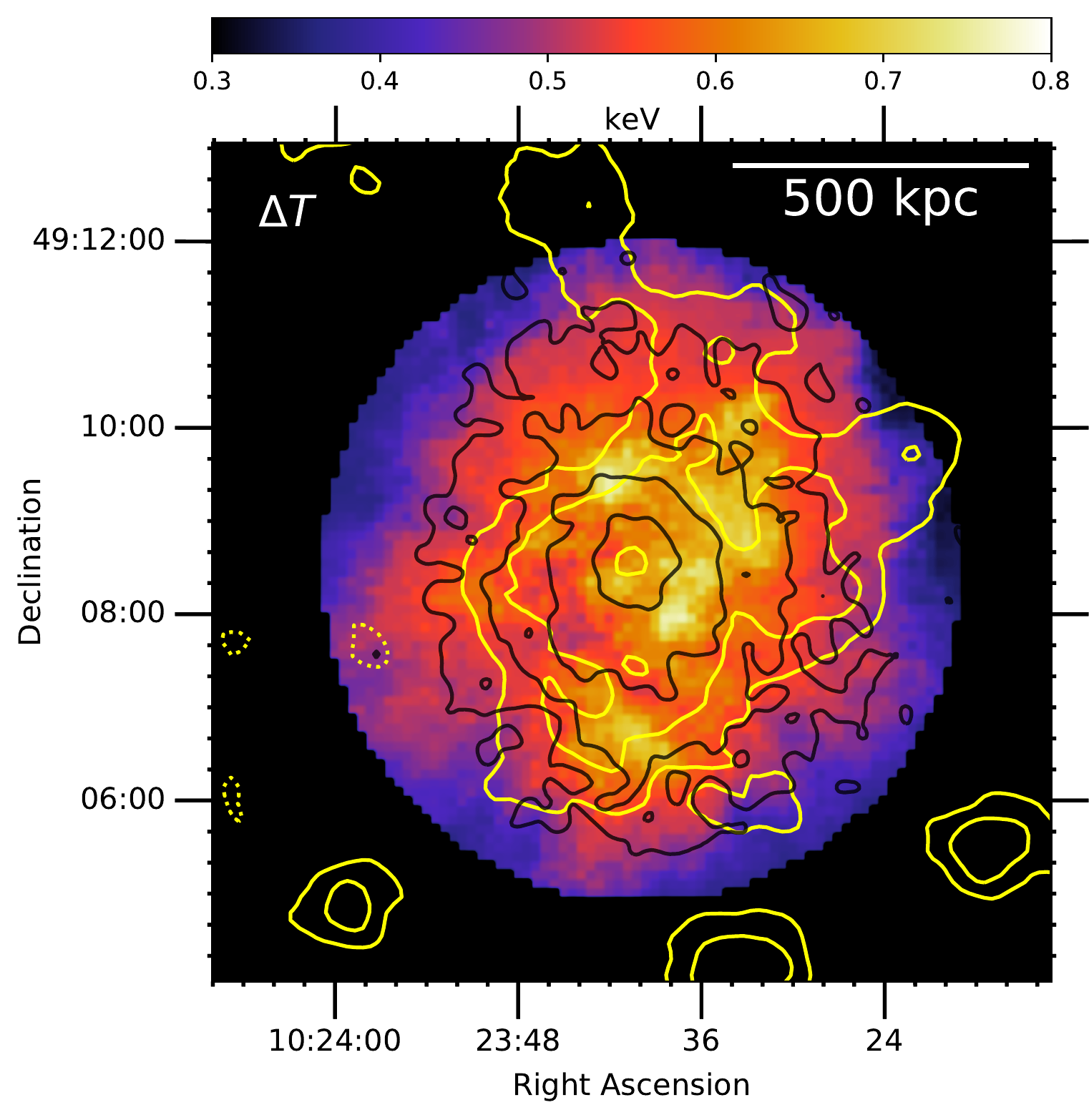}
	\caption{Temperature and error maps overlaid with the LOFAR low-resolution (yellow) and Chandra (black) contours. The contour levels are identical to those in Figure \ref{fig:a990_chandra}. The cluster has a mean temperature of 5~keV. }
	\label{fig:a990_T}
\end{figure*}

We use $\mathtt{ClusterPyXT}$ package to perform spectral fitting of the X-ray data and to produce temperature map in Figure \ref{fig:a990_T}. 
The temperature map shows that the ICM gas has a mean temperature of about $5\,{\rm keV}$ which is a moderate value for the known clusters \citep[e.g.][]{Frank2013,Zhu2016}. Across the cluster, the distribution of the ICM gas temperature is patchy. The temperature is slightly higher in the NW region (i.e.  $\sim$$5.4\,{\rm keV}$) than that in the SE region of the cluster (i.e. $\sim$$5.1\,{\rm keV}$). In the core region, the mean temperature is about $5.2\,{\rm keV}$ around the X-ray emission peak and slightly increases to $5.4\,{\rm keV}$ at the distance of 120~kpc. However, these variations are still within the mean errors of $\sim$$0.55\,{\rm keV}$ (see Figure \ref{fig:a990_T}, right). This is due to the short exposure duration of the X-ray observations (i.e. 10~ks). Future deep X-ray observations will provide information on the possible temperature variations in the cluster.

\section{Discussion}
\label{sec:disc}

We have reported the detection of an extended radio source in A990 and classify it as a radio halo. The classification is based on (i) its extended size of $700\,{\rm kpc}$ that is not associated with any compact sources, (ii) its location at the cluster centre, (iii) its extended emission largely overlaid the X-ray extended emission from the ICM. A990 with a mass of $M_{500}=(4.9\pm0.3)\times10^{14}\,M_\odot$ \citep[][]{Planck2015} is among the less massive clusters known to host a giant radio halo.

The power for radio halos is well known to correlate with the mass and X-ray luminosity of their host clusters \citep[e.g.][]{Cassano2013}. To examine the correlations in case of A990, we add our data points to the existing $P_{\rm 1.4~GHz}-M_{\rm 500}$ and $P_{\rm 1.4~GHz}-L_{500}$ correlations (see Figure \ref{fig:correlations}). Here $L_{500}$ is the X-ray luminosity measured within a radius $R_{500}$ at which the total particle density is 500 times the critical density of the Universe at the cluster redshift. If $\alpha=-1.0$, the radio power at 1.4~GHz is $P_{\rm 1.4~GHz}=1.2\pm0.1\times10^{23}\,{\rm W\,Hz}^{-1}$ which is well below the expected value for a cluster that has a mass of $(4.9\pm0.3)\times10^{14}\,M_\odot$ (i.e. 5 times less power than expected) and X-ray luminosity of $(3.66\pm0.08)\times10^{44}\,{\rm ergs\,s}^{-1}$ (i.e. 4 times underluminous in radio power). In case of $\alpha=-1.3$ (see Sec. \ref{sec:radio}), the halo power is lower, i.e. $P_{\rm 1.4~GHz}=6.1\times10^{22}\,{\rm W\,Hz^{-1}}$ and is significantly below the correlations (i.e. 9 and 8 times lower than expected from the $P_{\rm 1.4~GHz}-M_{\rm 500}$ and $P_{\rm 1.4~GHz}-L_{500}$  correlations, respectively) and the upper limit values for undetected halos (Figure \ref{fig:correlations}). The difference could be that A990 lies in the poorly-constrained regions for less-massive clusters in the $P_{\rm 1.4~GHz}-M_{\rm 500}$ and $P_{\rm 1.4~GHz}-L_{500}$ correlations. However, it is still interesting that the halo in A990 is also about 3 times less powerful than the typical upper limits for halos that are thought to have ultra-steep spectra (i.e. blue arrows in Figure \ref{fig:correlations}).
Our morphological analysis of the cluster A990 with a mass of $(4.9\pm0.3)\times10^{14}\,M_\odot$ and its position in the $P_{\rm 1.4~GHz}-M_{500}$ and $P_{\rm 1.4~GHz}-L_{500}$ diagrams suggest that A990 can possibly host a radio halo with an ultra-steep spectrum. However, adequate higher- or lower-frequency observations are necessary to constrain its spectrum and confirm the scenario.

In the turbulence re-acceleration model, radio halos are formed through merger induced turbulence and are likely to be detected in massive merging clusters. In these systems,  a fraction of an amount of turbulent energy flux is channelled into the acceleration of relativistic particles and the amplification/generation of magnetic fields in the ICM during cluster mergers. In less-massive clusters, merging events inject less energy into the ICM, resulting in radio halos being formed less frequently and with steeper spectra \cite[e.g.][]{Brunetti2008,Cassano2010a,Cassano2012}. The model predicts that only about $10-15$ percent of clusters with a mass below $6.5\times10^{14}\,{M_\odot}$ hosts Mpc-scale radio halos with spectra flatter than $\alpha$$\sim$$-1.5$  \citep[e.g.][]{Cuciti2015}. At these masses, giant halos with steeper spectra are expected to be more common and the fraction of these clusters with halos increase up to 40 percent including steeper spectrum halos \citep[e.g.][]{Cassano2006,Cassano2010a,Cuciti2015}. Therefore, the detection of radio halos in clusters with $M_{500}\lesssim6.5\times10^{14}\,M_\odot$ is expected to be more common at low frequencies with LOFAR \citep{VanHaarlem2013}, MWA \citep[Phase II;][]{Wayth2018}, and uGMRT \citep{Gupta2017}. Our discovery of the radio halo in A990 with LOFAR and the fact that it sits below the correlations are in line with such expectations.

The low power of A990 in the $P_{\rm 1.4~GHz}-M_{500}$ and $P_{\rm 1.4~GHz}-L_{500}$ planes could also indicate a particular phase in the evolutionary state of the cluster. Magnetohydrodynamic (MHD) simulations of the re-acceleration of cosmic ray electrons by merger induced turbulence by \cite{Donnert2013} modelled the evolution of a radio halo power during cluster mergers. The radio halo becomes brighter after the merger and it is fainter in a later merging stage. At the same time the halo spectrum is flattened in the early stage of the merger, but is steepened below $-1.5$ after a few Gyrs of core passage. The halo could be observed for a few Gyrs depending on the observing frequencies which probe different energy of the population of relativistic electrons and magnetic field strength. At low frequencies, the halo is observable much longer due to the large lifetime of  relativistic electrons with lower energy and weak magnetic fields. 

In general, spectral information are thus crucial to understand the origin and evolution of radio halos. On-going radio surveys of a large fraction of the sky such as LoTSS \citep[][]{Shimwell2017,Shimwell2019} will provide statistical estimates on the fraction of radio halos in complete samples of clusters down to relatively small masses. In combination with high-frequency observations with the ASKAP/EMU (e.g. clusters with declination between 0 and $+30^{\circ}$), uGMRT and VLA telescopes, the halo spectra will be measured which will improve our understanding of the mechanism of particle acceleration and magnetic field amplification/generation at cluster scales. 

\section{Conclusions}

In this paper we present LOFAR 144~MHz observations, as part of the LoTSS surveys, of the galaxy cluster Abell 990 ($z=0.144$) with a mass of $M_{500}=(4.9\pm0.3)\times10^{14}\,M_\odot$. We detect an extended radio emission with a projected size of $\sim$$700\,{\rm kpc}$ at the cluster centre. Due to its location and size, we classify the extended radio emission as a radio halo.

The flux density at 144~MHz for the halo is $S_{\rm 144~MHz}=20.2\pm2.2\,{\rm mJy}$ (i.e. above $3\sigma$ contours),  corresponding to a radio power of  $P_{\rm 144~MHz}=(1.2\pm0.1)\times10^{24}~{\rm W~Hz}^{-1}$ ($k$-corrected). The radio halo is one of the two lowest power radio halos detected to date.

Our analysis on the dynamical status of the cluster using the X-ray data reveals that the cluster has a slightly disturbed ICM and does not have major merger activities. The X-ray luminosity of the halo between 0.1--2.4~keV is  $L_{\rm 500} = (3.66\pm0.08)\times10^{44}\,{\rm ergs\,s}^{-1}$. The cluster mean temperature is relatively low, at $5\,{\rm keV}$. There are possible variations in the temperature, but require deep X-ray observations to confirm.

Our measurements on the radio power, X-ray luminosity, and the ICM temperature are at the low ends. This can be explained by the cluster being a dynamically disturbed, less massive system that contains less energy than more massive clusters. 

Our study reveals the potential of detecting many faint radio halos in  less-massive galaxy clusters with LOFAR. With the full coverage of the northern hemisphere, the LoTSS survey will provide the fraction of clusters of this mass hosting radio halos that will be compared with the theoretical predictions \citep[e.g.][]{Cassano2010a}. A combination with high-frequency observations will provide spectral information on the sources that are important for studies of the origin of radio halos.

\section*{Acknowledgements}
We thank the anonymous referee for the helpful comments. 
DNH and A.Bonafede acknowledge support from the ERC-Stg17 714245 DRANOEL. 
MB acknowledges support from the Deutsche Forschungsgemeinschaft under Germany's Excellence Strategy - EXC 2121 "Quantum Universe" - 390833306. 
A.Botteon, EO, and RJvW  acknowledge support from the VIDI
research programme with project number 639.042.729, which is
financed by the Netherlands Organisation for Scientific Research
(NWO).
HR acknowledges support from the ERC Advanced Investigator programme NewClusters 321271. 
AD acknowledges support by the BMBF Verbundforschung under the grant 05A17STA : The Jülich LOFAR Long Term Archive and the German LOFAR network are both coordinated and operated by the Jülich Supercomputing Centre (JSC), and computing resources on the supercomputer JUWELS at JSC were provided by the Gauss Centre for supercomputing e.V. (grant CHTB00) through the John von Neumann Institute for Computing (NIC).
These data were (partly) processed by the LOFAR Two-Metre Sky Survey (LoTSS) team. This team made use of the LOFAR direction independent calibration pipeline (\url{https://github.com/lofar-astron/prefactor}) which was deployed by the LOFAR e-infragroup on the Dutch National Grid infrastructure with support of the SURF Co-operative through grants e-infra 160022 e-infra 160152 (Mechev et al. 2017, PoS(ISGC2017)002). The LoTSS direction dependent calibration and imaging pipeline (\url{http://github.com/mhardcastle/ddf-pipeline}) was run on compute clusters at Leiden Observatory and the University of Hertfordshire which are supported by a European Research Council Advanced Grant [NEWCLUSTERS-321271] and the UK Science and Technology Funding Council [ST/P000096/1].
We have including the lotss-dr2-infrastructure list as this project made use of the DR2 pipeline and post processing extraction/selfcal scheme.
The scientific results reported in this article are based  on data obtained from the Chandra Data Archive.  This research has made use of software provided by the Chandra X-ray Center (CXC) in the application packages CIAO, ChIPS, and Sherpa.
The National Radio Astronomy Observatory is a facility of the National Science Foundation operated under cooperative agreement by Associated Universities, Inc.
 %
\section*{Data Availability}
The data underlying this article will be shared on reasonable request to the corresponding author.



\bibliographystyle{mnras}
%




\appendix
\section{Optical counterparts of compact sources}
\begin{figure*}
	\centering
	\includegraphics[width=0.275\textwidth]{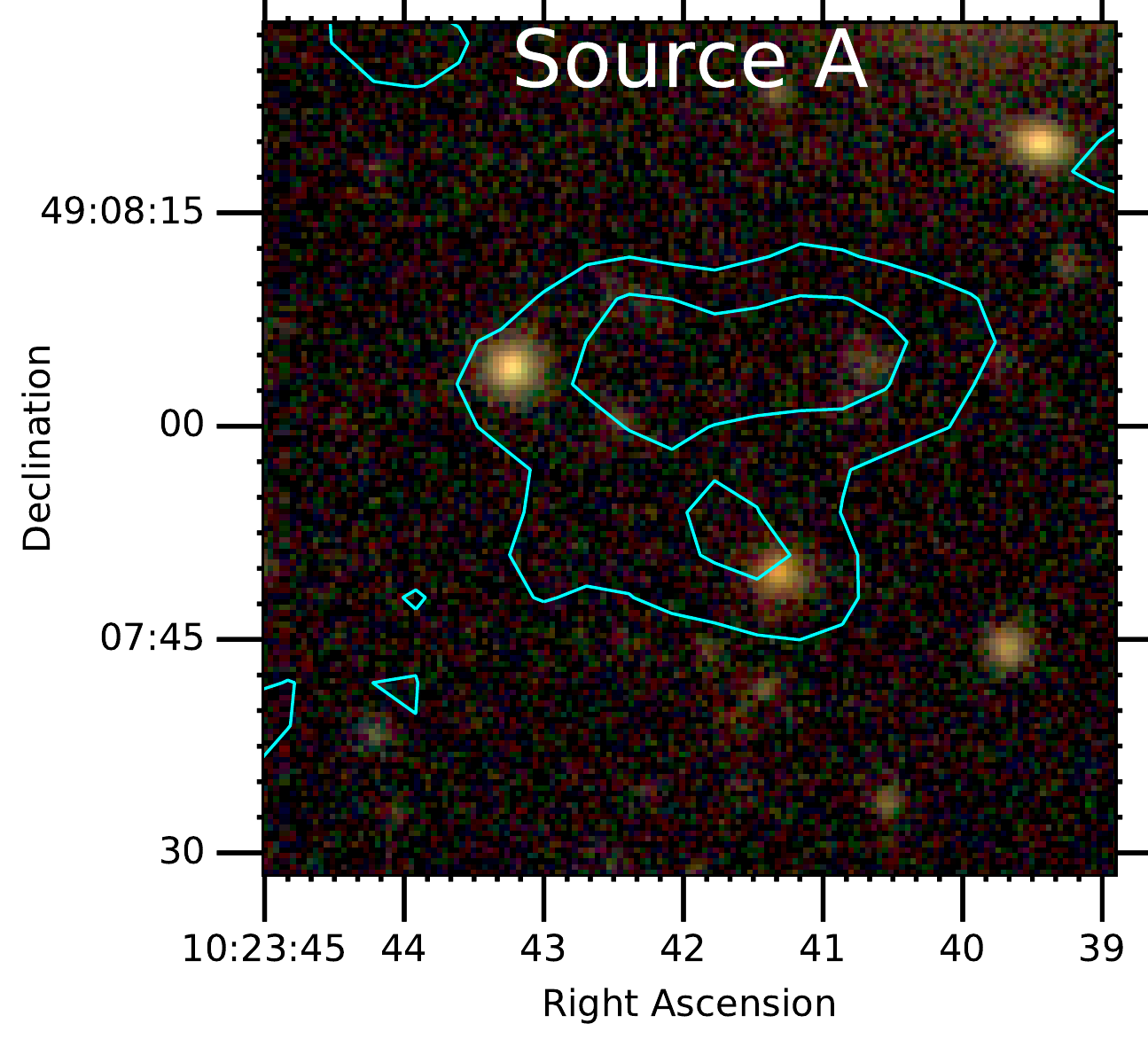} \hfill
	\includegraphics[width=0.275\textwidth]{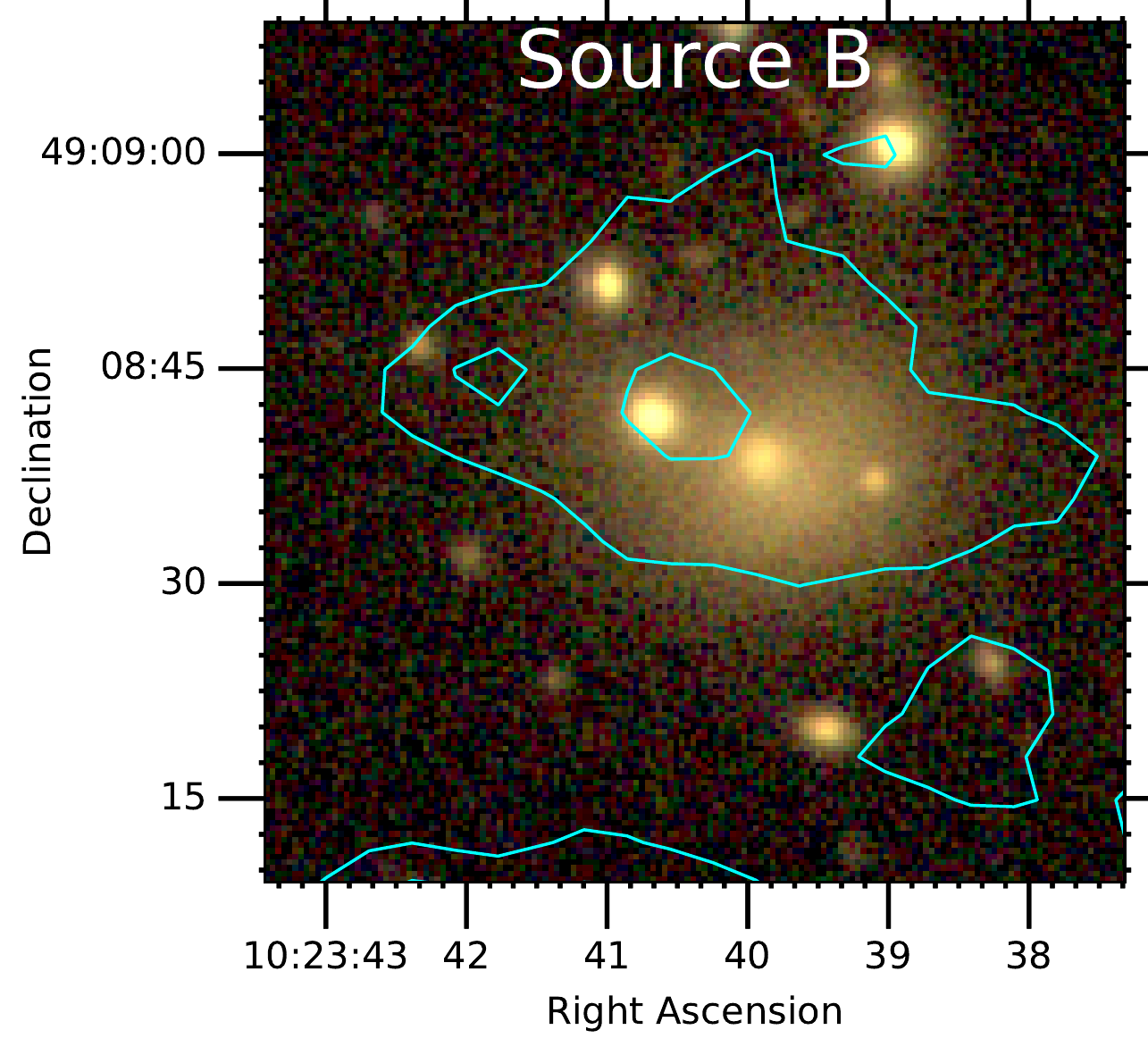} \hfill
	\includegraphics[width=0.275\textwidth]{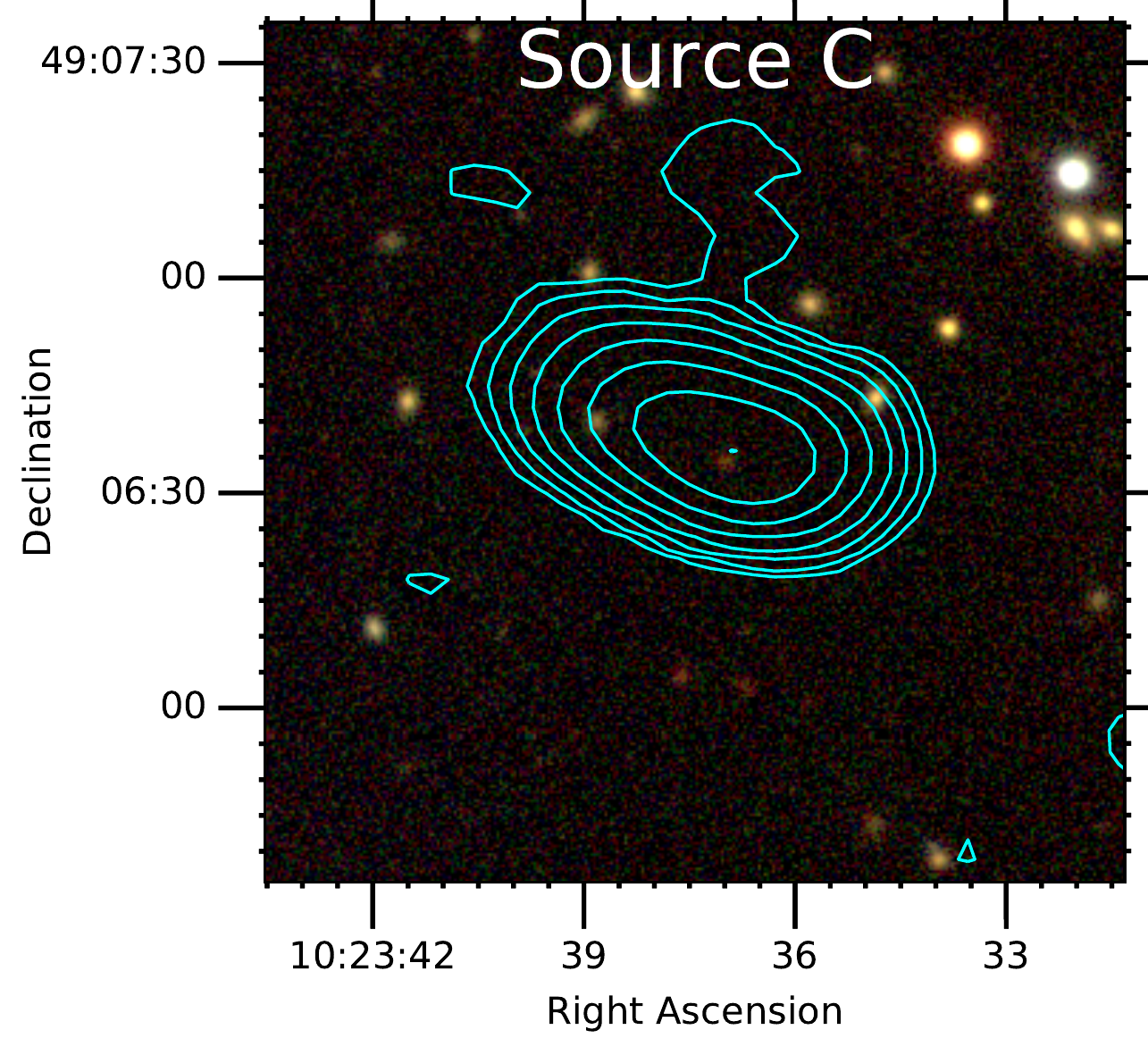} \\
	\includegraphics[width=0.275\textwidth]{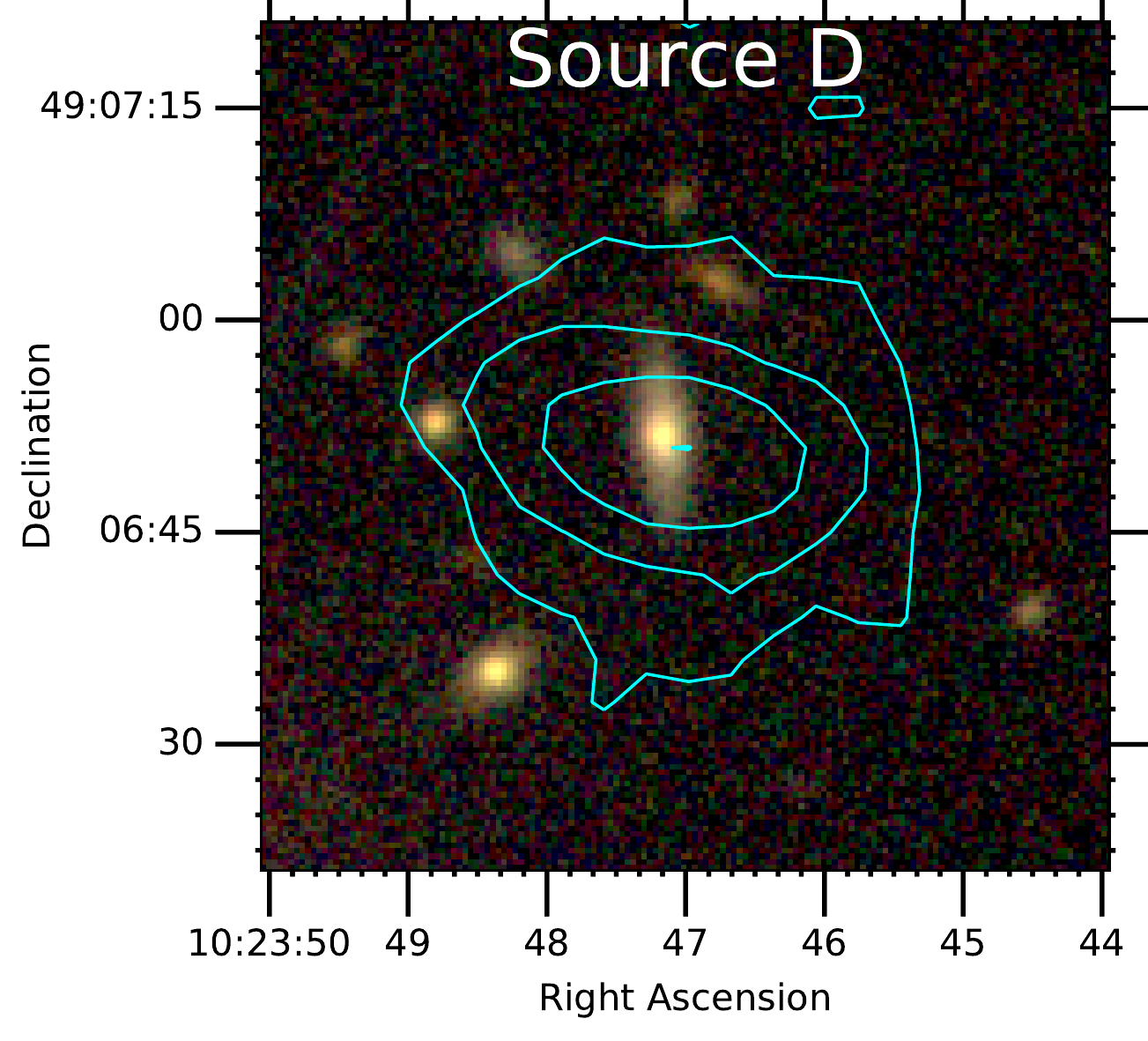} \hfill
	\includegraphics[width=0.275\textwidth]{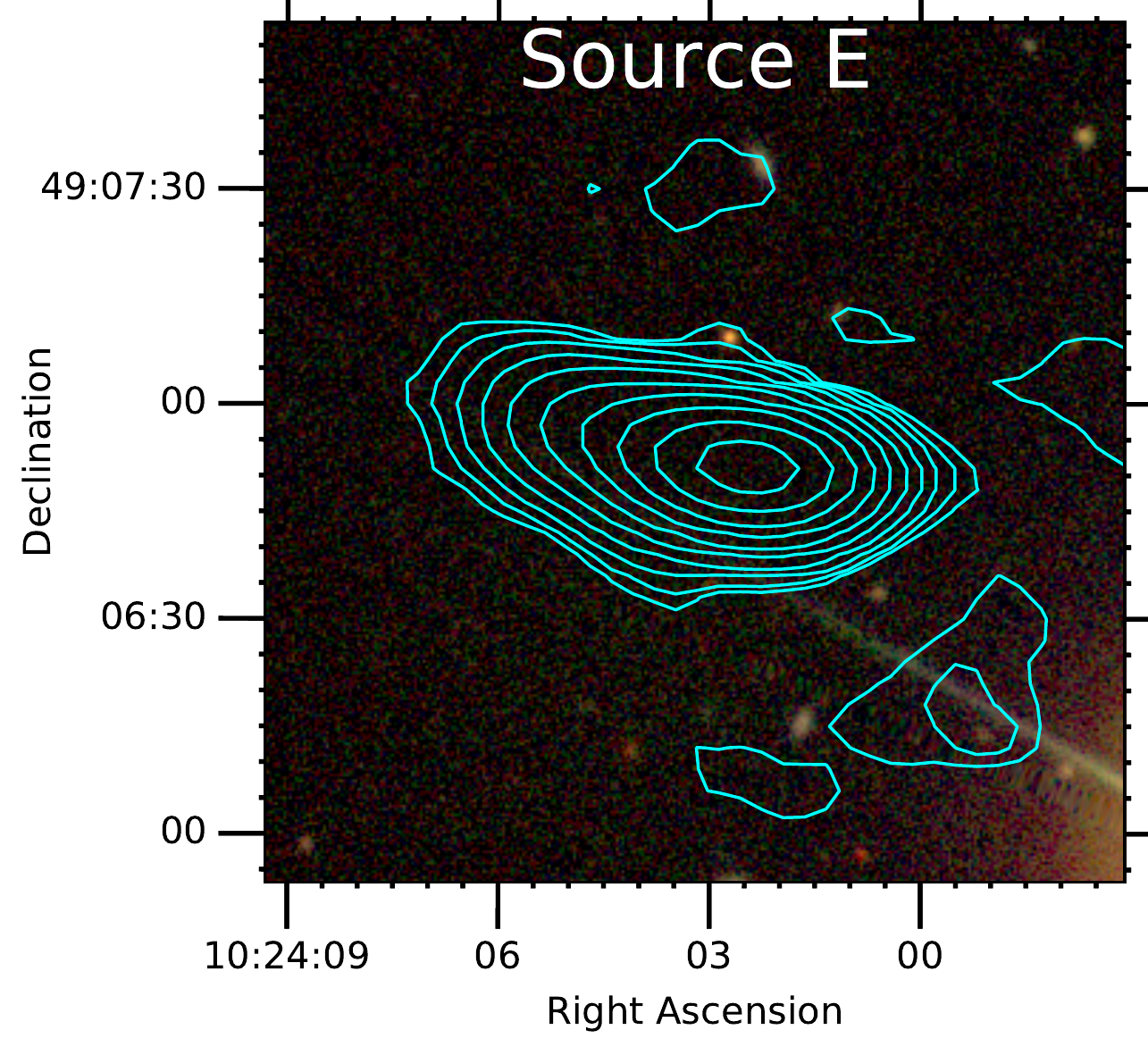} \hfill
	\includegraphics[width=0.275\textwidth]{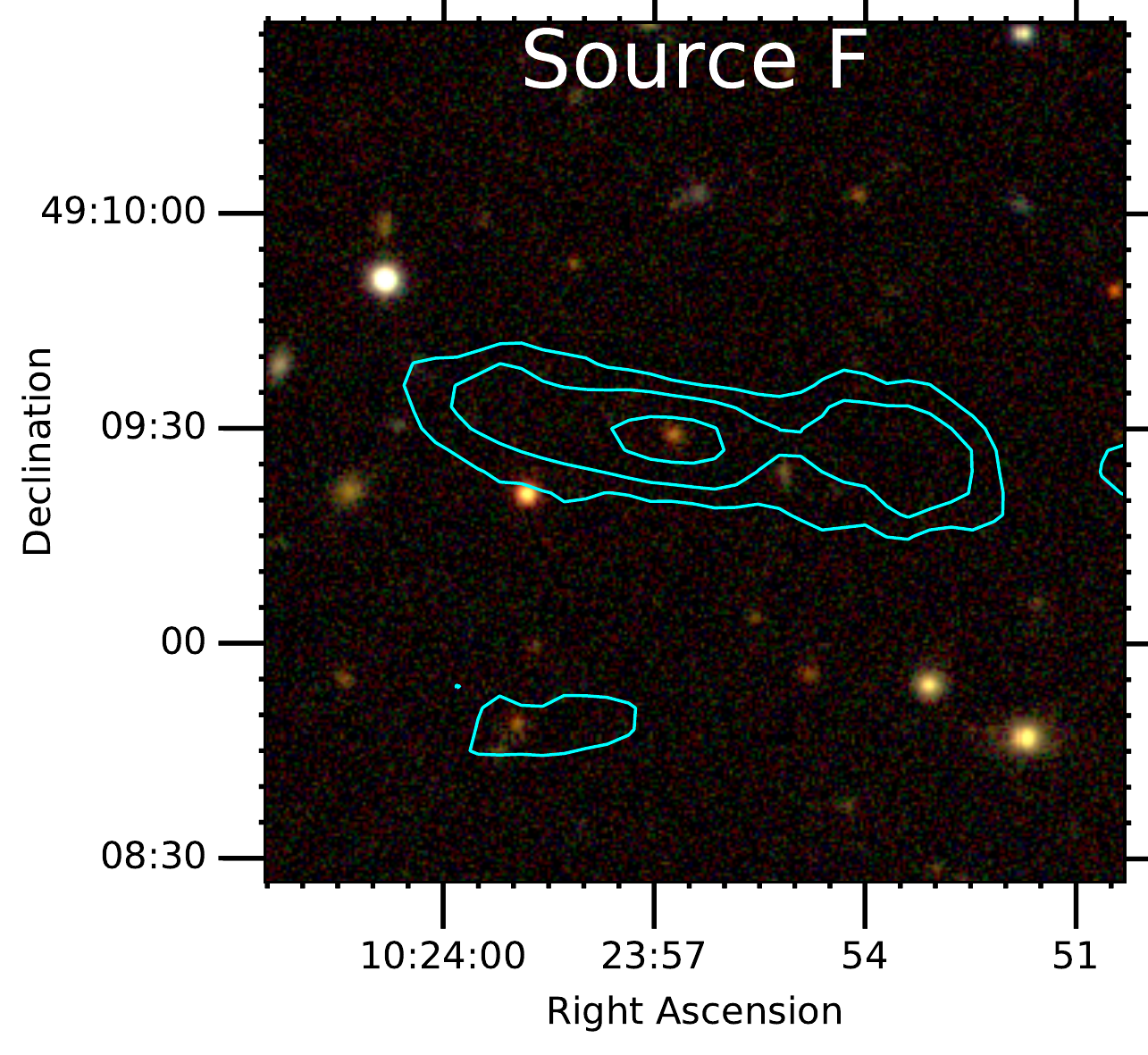} \\
	\includegraphics[width=0.275\textwidth]{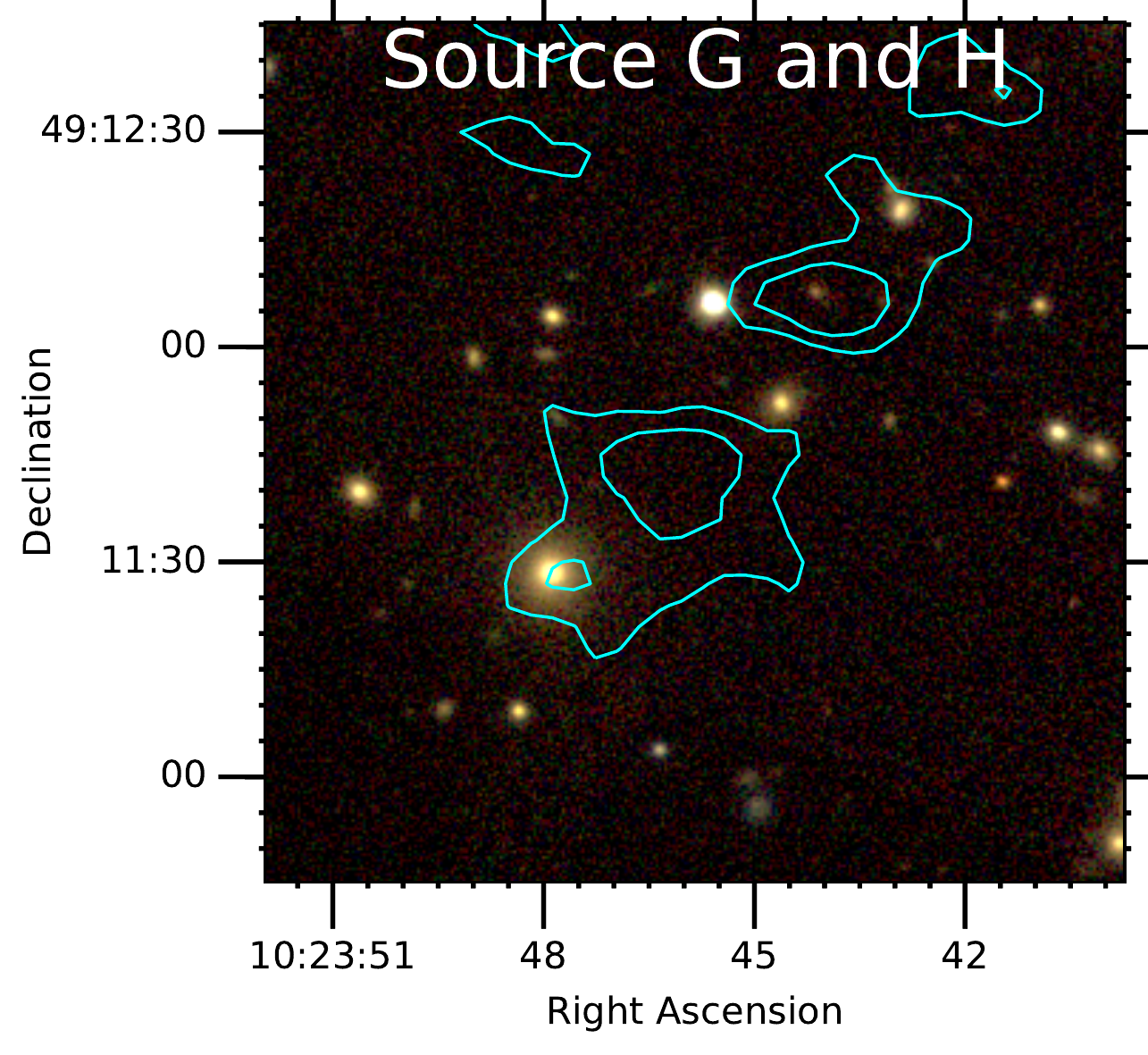} \hfill
	\includegraphics[width=0.275\textwidth]{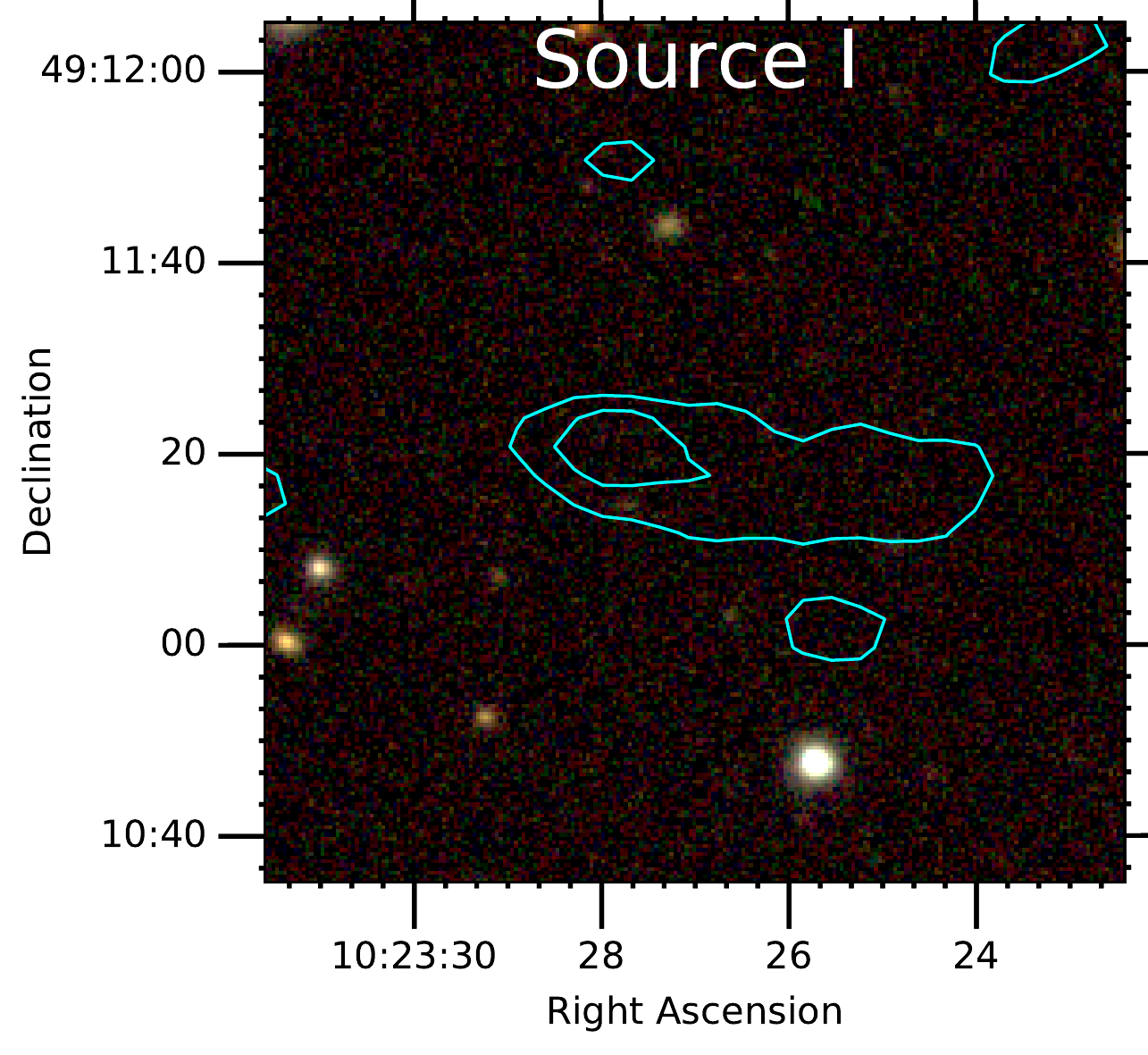} \hfill
	\includegraphics[width=0.275\textwidth]{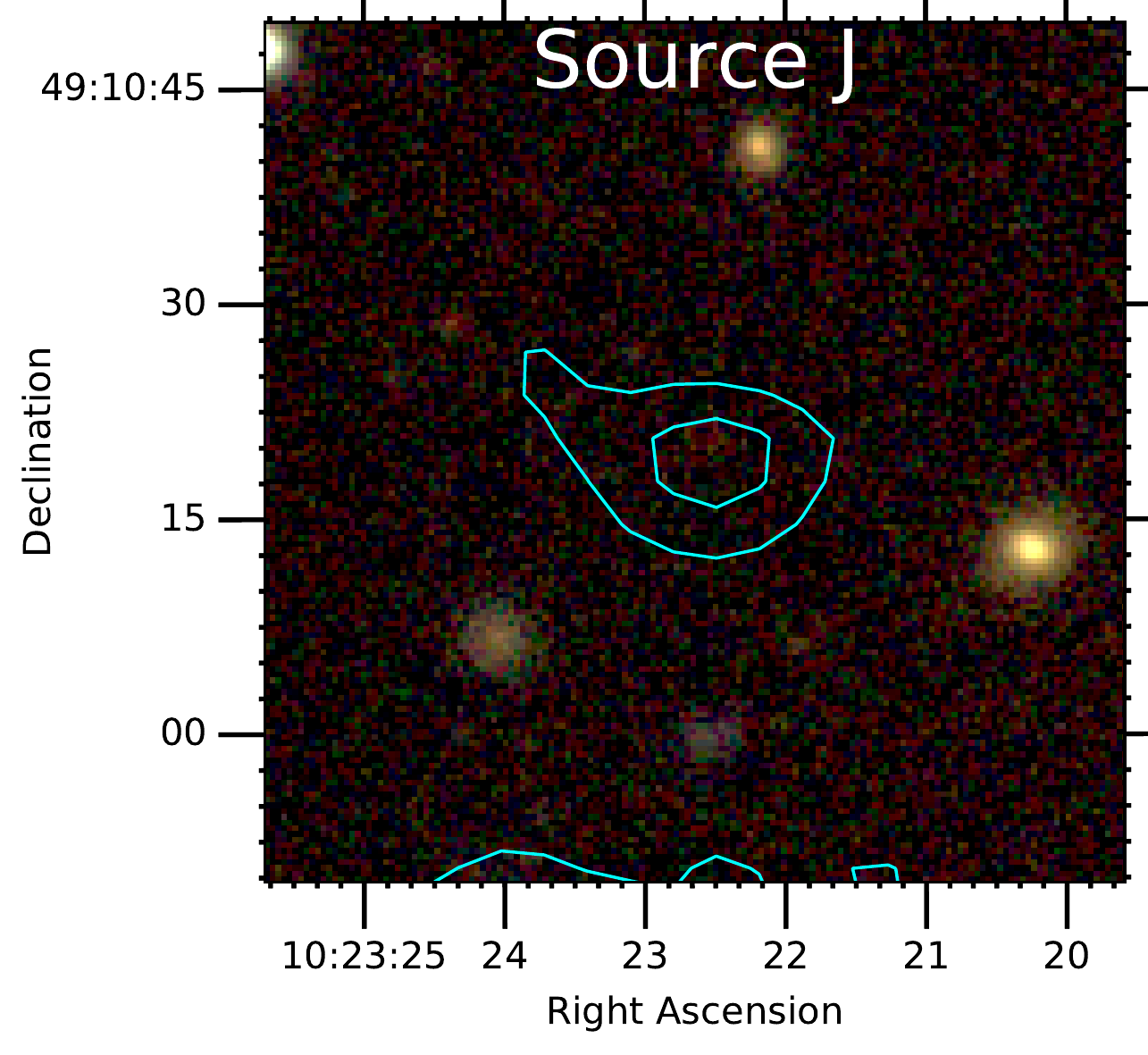} \\
	\includegraphics[width=0.275\textwidth]{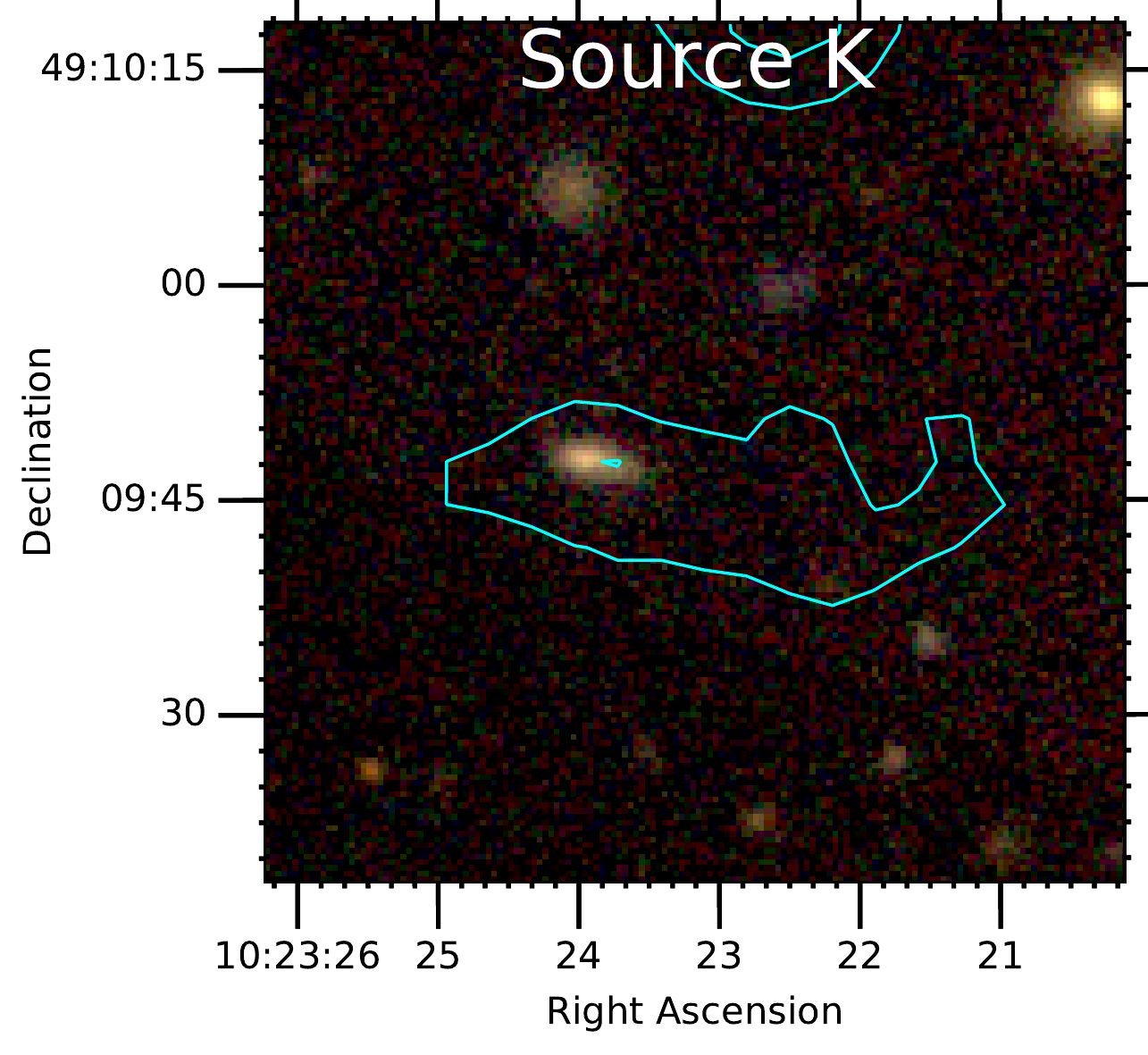} \hfill
	\includegraphics[width=0.275\textwidth]{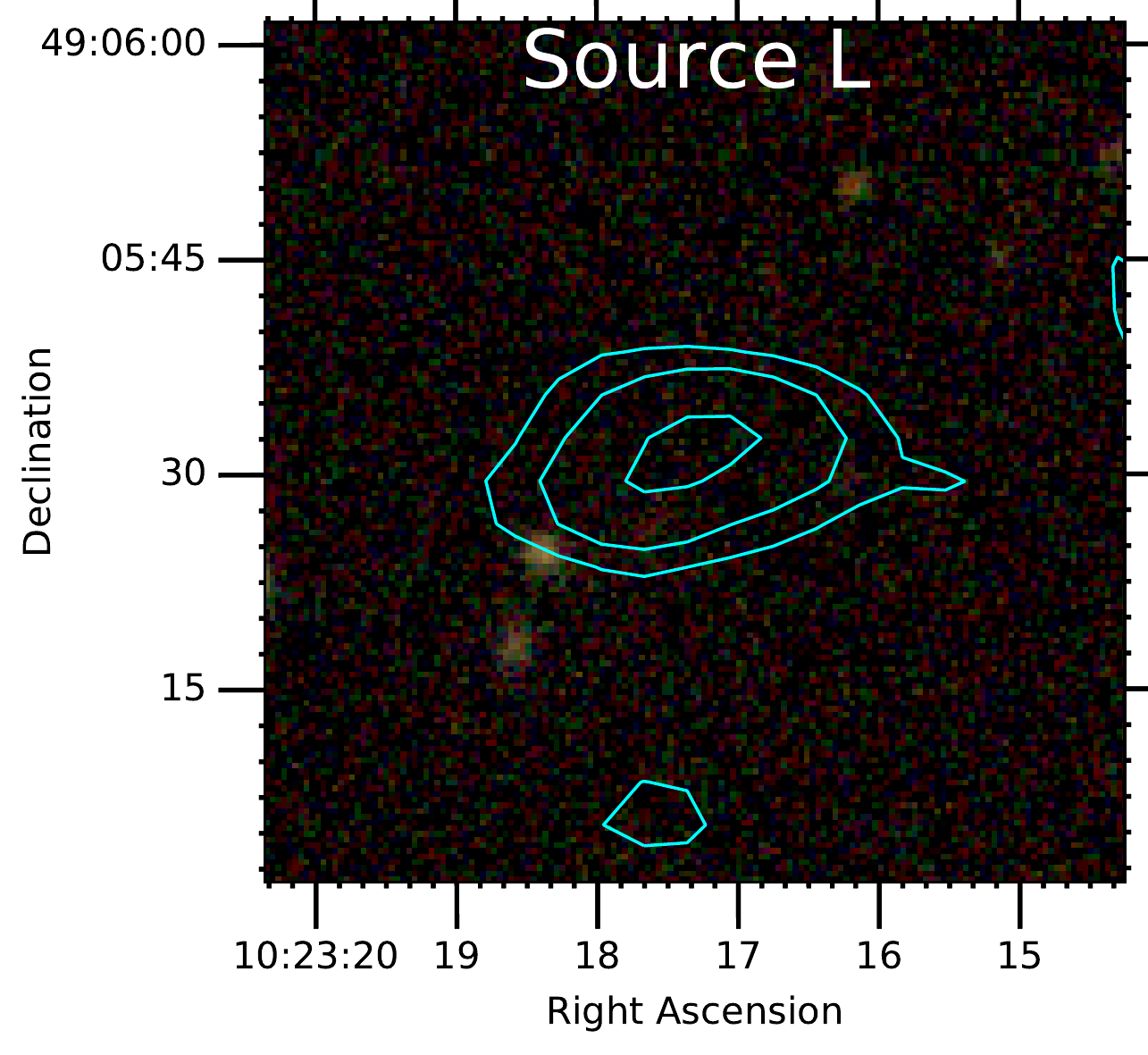} \hfill
	\includegraphics[width=0.275\textwidth]{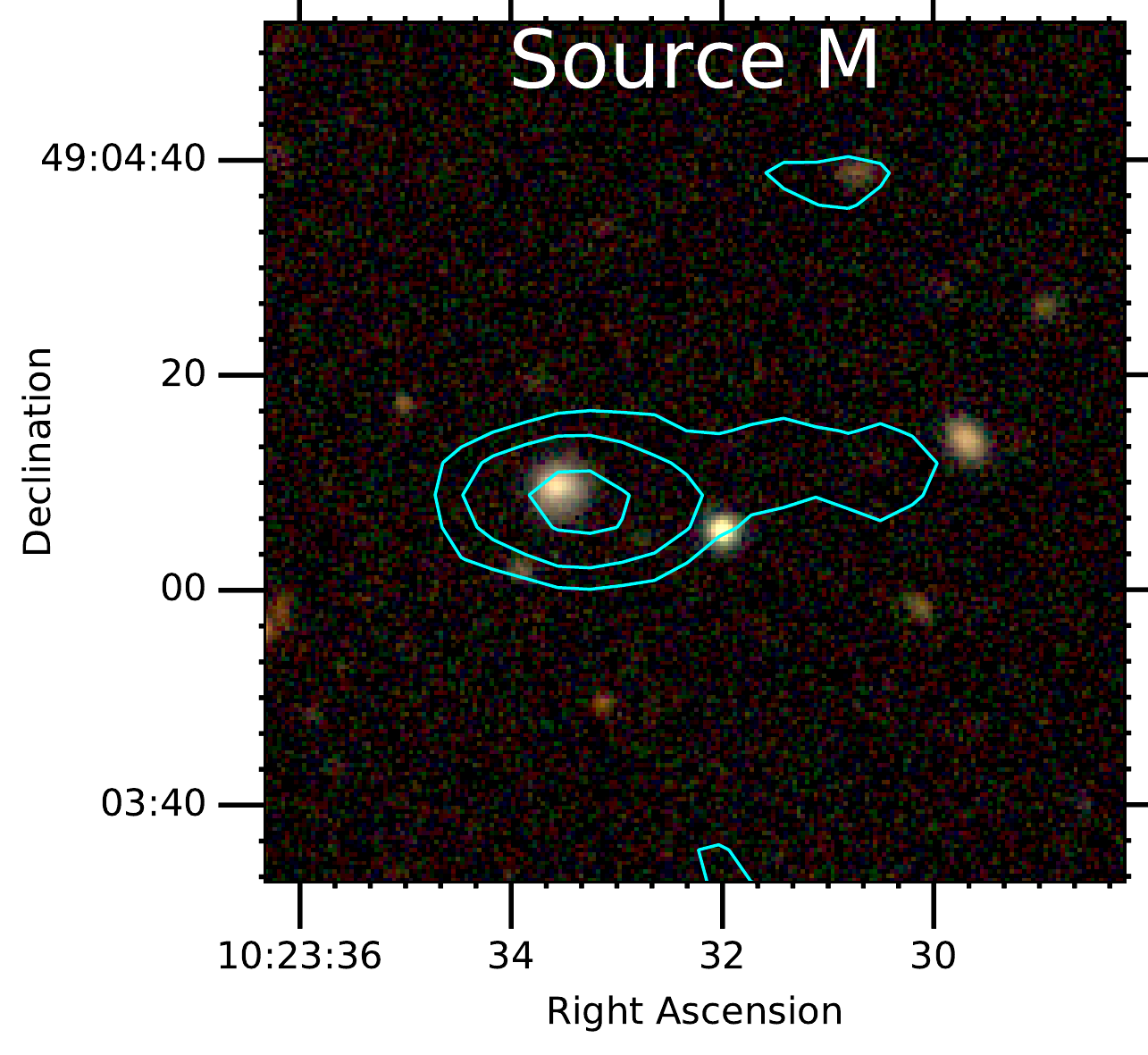} \\
	\includegraphics[width=0.275\textwidth]{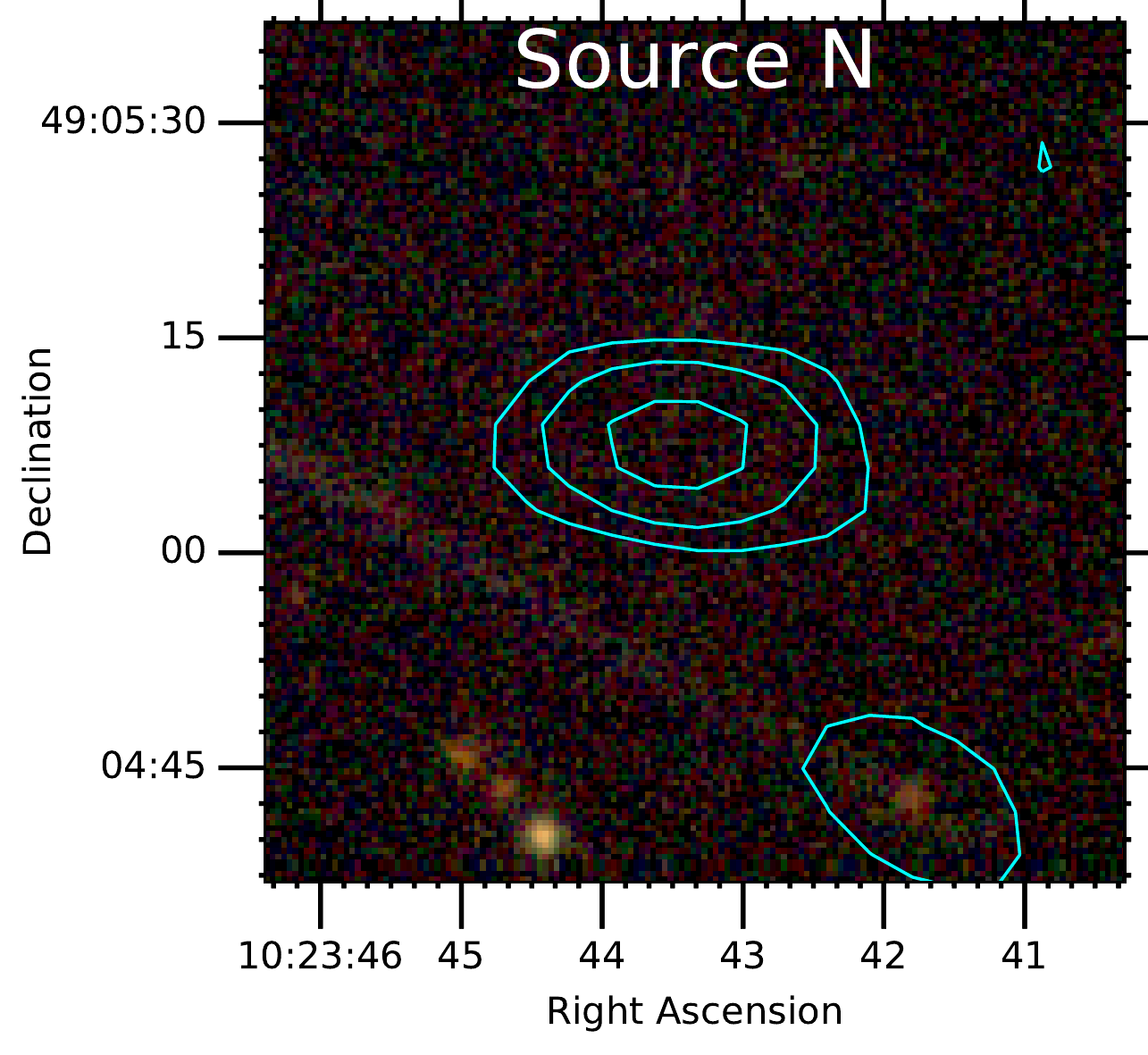} \hfill
	\includegraphics[width=0.275\textwidth]{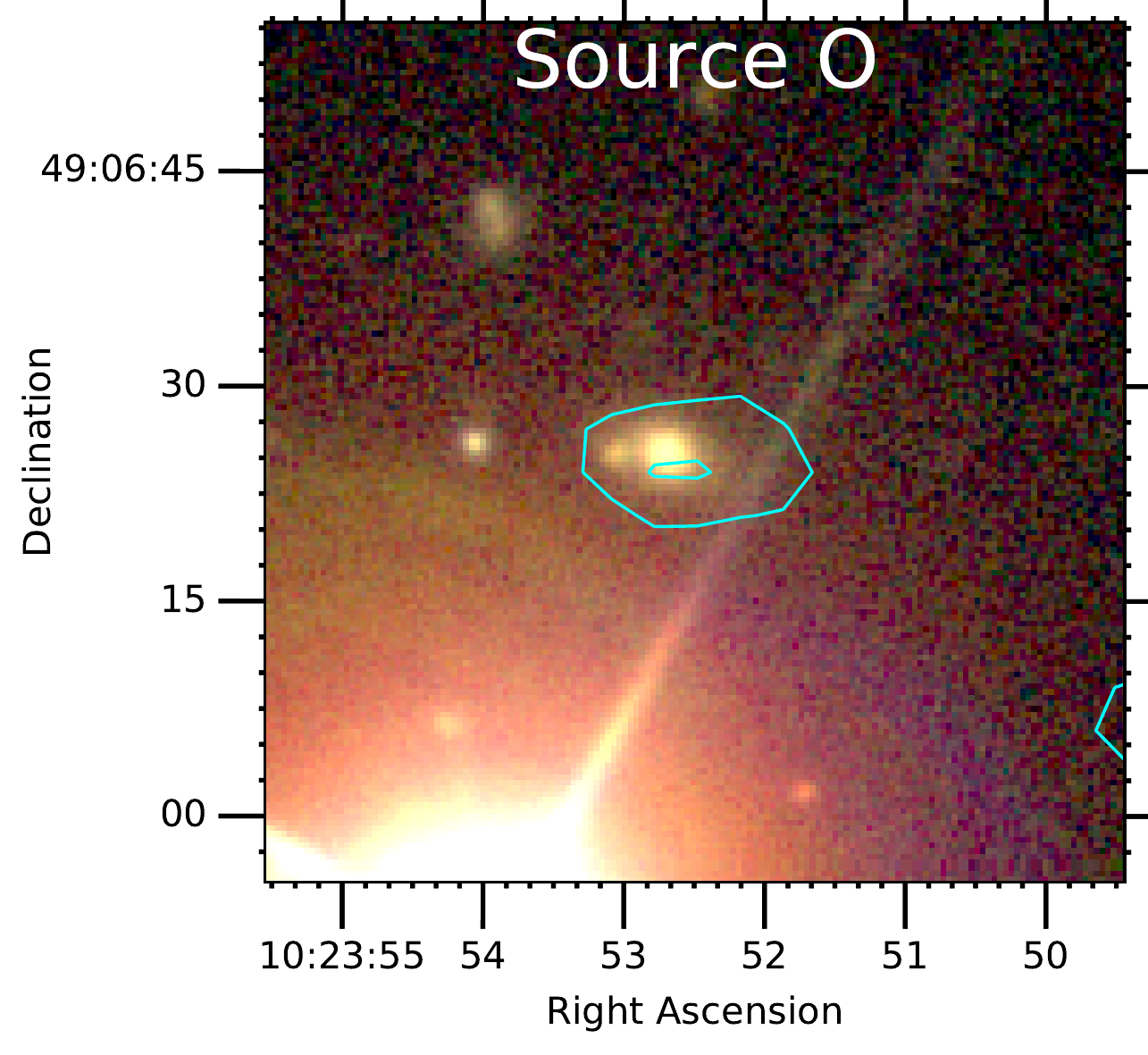} \hfill
	\includegraphics[width=0.275\textwidth]{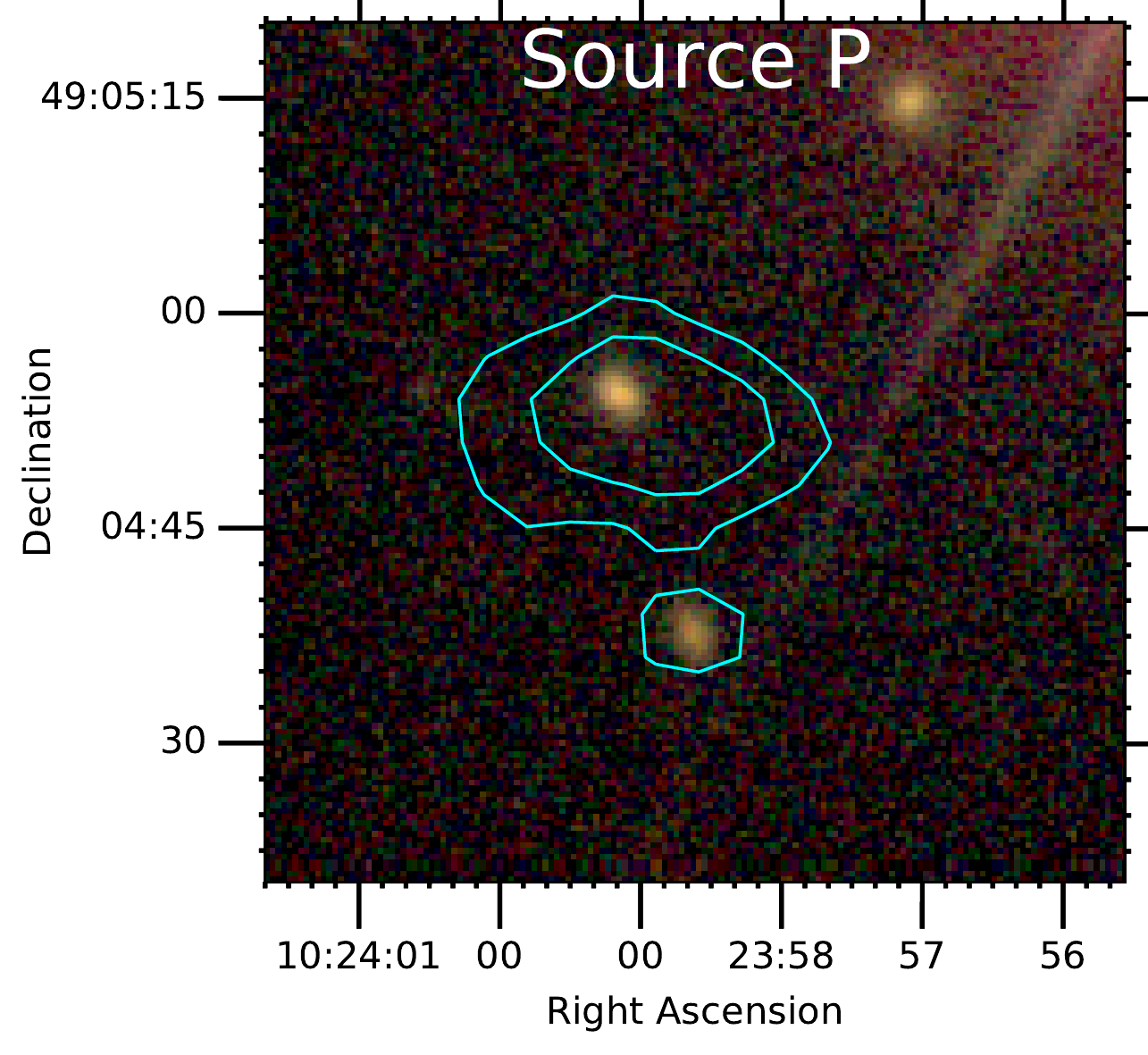} \\
	 
	\caption{SDSS optical (i, r, and g band) images overlaid with the LOFAR high-resolution contours. The contours are at the same levels as those in Figure \ref{fig:a990_lofar}. The source labels are given in Figure \ref{fig:a990_label}.}
	\label{fig:a990_cutouts}
\end{figure*}

%
To identify optical counterparts of the radio compact sources, we overlaid the LOFAR high-resolution contours on SDSS  images in Figure \ref{fig:a990_cutouts}.

\bsp	
\label{lastpage}
\end{document}